\documentclass[aps,prb,twocolumn,groupedaddress,floatfix]{revtex4}
\usepackage{amsmath}
\usepackage{amssymb}
\usepackage{graphicx}
\usepackage{bm}

\newlength{\figurewidth} 
\newtheorem{theorem}{Theorem}

\newtheorem{proposition}[theorem]{Proposition}

\input{tcilatex}
\begin{document}

\title{Langevin Thermostat for Rigid Body Dynamics}
\author{Ruslan L. Davidchack}
\author{Richard Handel}
\author{M.V. Tretyakov}
\affiliation{Department of Mathematics, University of Leicester, Leicester, LE1 7RH,
United Kingdom}

\begin{abstract}
We present a new method for isothermal rigid body simulations using the
quaternion representation and Langevin dynamics. It can be combined with the
traditional Langevin or gradient (Brownian) dynamics for the translational
degrees of freedom to correctly sample the $NVT$ distribution in a
simulation of rigid molecules. We propose simple, quasi-symplectic
second-order numerical integrators and test their performance on the TIP4P
model of water. We also investigate the optimal choice of thermostat
parameters.
\end{abstract}

\maketitle


\section{Introduction}

Classical molecular dynamics simulation of an isolated system naturally
samples states from a microcanonical ($NVE$) ensemble, where the number of
particles $N$, volume $V$, and total energy of the system $E$ are held
constant. However, in many cases it is desirable to study the system in a
more experimentally relevant canonical ($NVT$) ensemble, where the
temperature $T$ is specified instead of $E$. In order to sample from the
canonical ensemble, the molecular dynamics equations of motion are modified
by introducing the interaction of the system with a \textquotedblleft
thermostat\textquotedblright . There exist a large variety of approaches for
introducing such a thermostat, which can be roughly classified into two
categories: deterministic and stochastic, depending on whether the resulting
equations of motion contain a random component (for a review, see, e.g. Ref.~%
\onlinecite{Hunenberger05}).






Among various deterministic approaches, those based on coupling the system
to a few external degrees of freedom (e.g. Nos\'{e}-Hoover thermostat) have
become very popular. Given ergodicity in the molecular system dynamics, such
thermostats are proven to generate correct canonical ensemble sampling of
the system phase space. However, since the thermostat variables are coupled
and control directly only global system quantities (e.g. kinetic energy),
such thermostats rely on the efficient energy transfer within the system to
achieve \emph{equipartition} within the canonical distribution, such that
the average energy of each degree of freedom within the system is equal to $%
k_{B}T$. Therefore, in a system where the energy transfer between its
different parts is slow (e.g., systems combining fast and slow degrees of
freedom), the simple Nos\'{e}-Hoover thermostat may have difficulty
maintaining the same temperature for the different parts of the system. In
this case more complicated thermostats are necessary, for example, Nos\'{e}%
-Hoover chain thermostat, or separate thermostats for different parts of the
systems (see, e.g. Ref.~\onlinecite{Ben}).

The stochastic approach exploits ergodic stochastic differential equations
(SDEs) with the Gibbsian (canonical ensemble) invariant measure. For this
purpose, Langevin-type equations or gradient systems with noise can be used
(see, e.g. Refs.~\onlinecite{SKE99,ISK01,Soize,Stua,MT7} and references
therein). Stochastic thermostats, with their independent thermalization of
each degree of freedom, provide direct control of equipartition and thus do
not need to rely on the efficient energy transfer within the system.


In order to achieve such a direct thermalization of the system, one needs to
be able to apply stochastic thermostats to all types of degrees of freedom.
The standard Langevin equations for translational degrees of freedom are
well known, while Langevin thermostats for systems with constraints have
been proposed quite recently~\cite{Eijnden06,Sun08}. In this paper we
introduce Langevin equations for the rigid body dynamics in the \emph{%
quaternion representation} and propose effective second-order
quasi-symplectic numerical integrators for their simulation. These equations
can be coupled either with Langevin or Brownian dynamics for the
translational degrees of freedom.


In Section~\ref{sec_mot} we recall the Hamiltonian system for rigid body
dynamics in the quaternion representation from Ref.~\onlinecite{qua02},
based on which we derive Langevin and gradient-Langevin thermostats.
Second-order (in the weak sense) numerical methods for these stochastic
systems are constructed in Section~\ref{sec_num}. We test the thermostats
and the proposed numerical integrators on the TIP4P model of water\cite%
{Jorgensen83}. The results of our numerical experiments are presented in
Section~\ref{sec_tes}. In particular, we investigate the optimal choice of
thermostat parameters and the discretization error of the numerical methods.
A summary of the obtained results is given in Section~\ref{sec_sum}.

\section{Equations of Motion\label{sec_mot}}

We consider a system of $n$ rigid three-dimensional molecules described by
the center-of-mass coordinates $\mathbf{r}=(r^{1\mathsf{T}},\ldots
,r^{n\,\mathsf{T}})^{\mathsf{T}}\in \mathbb{R}^{3n},$ $%
r^{j}=(r_{1}^{j},r_{2}^{j},r_{3}^{j})^{\mathsf{T}}\in \mathbb{R}^{3},$ and
the rotational coordinates in the quaternion representation $\mathbf{q}%
=(q^{1\,\mathsf{T}},\ldots ,q^{n\,\mathsf{T}})^{\mathsf{T}}\in
\mathbb{R}^{4n}$, $q^{j}=(q_{0}^{j},q_{1}^{j},q_{2}^{j},q_{3}^{j})^{\mathsf{T%
}}\in \mathbb{R}^{4},$ such that $|q^{j}|=1$.  We use standard matrix notations,
and ``${\mathsf T}$'' denotes transpose.
Following Ref.~%
\onlinecite{qua02}, we write the system Hamiltonian in the form
\begin{equation}
H(\mathbf{r},\mathbf{p},\mathbf{q},\bm{\pi})=\frac{\mathbf{p}^{\mathsf{T}}%
\mathbf{p}}{2m}+\sum_{j=1}^{n}\sum_{l=1}^{3}V_{l}(q^{j},\pi ^{j})+U(\mathbf{r%
},\mathbf{q}),  \label{a1}
\end{equation}%
where $\mathbf{p=}(p^{1\,\mathsf{T}},\ldots ,p^{n\,\mathsf{T}})^{%
\mathsf{T}}\in \mathbb{R}^{3n}$, $p^{j}=(p_{1}^{j},p_{2}^{j},p_{3}^{j})^{%
\mathsf{T}}\in \mathbb{R}^{3},$ are the center-of-mass momenta conjugate to $%
\mathbf{r}$, $\bm{\pi}=(\pi ^{1\,\mathsf{T}},\ldots ,\pi ^{n\,\mathsf{T}}%
)^{\mathsf{T}}\in \mathbb{R}^{4n}$, $\pi ^{j}=(\pi _{0}^{j},\pi
_{1}^{j},\pi _{2}^{j},\pi _{3}^{j})^{\mathsf{T}}\in \mathbb{R}^{4},$ are the
angular momenta conjugate to $\mathbf{q}$, and $U(\mathbf{r},\mathbf{q})$ is
the potential interaction energy. The second term represents the rotational
kinetic energy of the system with
\begin{equation}
V_{l}(q,\pi )=\frac{1}{8I_{l}}\left[ \pi ^{\mathsf{T}}S_{l}q\right]^{2},\ \
q,\pi \in \mathbb{R}^{4},\ \ l=1,2,3,  \label{a2}
\end{equation}%
where the three constant $4$-by-$4$ matrices $S_{l}$ are such that
\begin{eqnarray*}
S_{1}q=(-q_{1},q_{0},q_{3},-q_{2})^{\mathsf{T}},
\!\!&&\!\!S_{2}q=(-q_{2},-q_{3},q_{0},q_{1})^{\mathsf{T}}, \\
S_{3}q &=&(-q_{3},q_{2},-q_{1},q_{0})^{\mathsf{T}},
\end{eqnarray*}%
and $I_{l}$ are the principal moments of inertia of the rigid molecule. The
Hamilton equations of motion are
\begin{equation}  \label{a20}
\begin{split}
\frac{d\mathbf{r}}{dt} &=\frac{\mathbf{p}}{m}\,,\quad \frac{d\mathbf{p}}{dt}
=-\nabla _{\mathbf{r}}U(\mathbf{r},\mathbf{q})\,, \\
\frac{dq^{j}}{dt} &=\sum_{l=1}^{3}\nabla _{\pi ^{j}}V_{l}(q^{j},\pi ^{j})\,,
\\
\frac{d\pi ^{j}}{dt} &=-\sum_{l=1}^{3}\nabla _{q^{j}}V_{l}(q^{j},\pi
^{j})-\nabla _{q^{j}}U(\mathbf{r},\mathbf{q})\,, \\
j&=1,\ldots ,n\,.
\end{split}%
\end{equation}
It is easy to check that if the initial conditions are chosen such that $%
|q^{j}(0)|=1$, then the corresponding Hamilton equations of motion ensure
that
\begin{equation}
|q^{j}(t)|=1\,,\ \ j=1,\ldots ,n\,,\ \ \mbox{for all~}t\geq 0.  \label{a21}
\end{equation}

In the rest of this section we derive stochastic thermostats for this
molecular system, which preserve (\ref{a21}). They take the form of ergodic
SDEs with the Gibbsian (canonical ensemble) invariant measure possessing the
density
\begin{equation}
\rho (\mathbf{r},\mathbf{p},\mathbf{q},\bm{\pi})\varpropto \exp (-\beta H(%
\mathbf{r},\mathbf{p},\mathbf{q},\bm{\pi })),  \label{a3}
\end{equation}%
where $\beta =1/(k_{B}T)>0$ is an inverse temperature.


\subsection{Langevin-type equations\label{sec_lan}}

Consider the Langevin-type equations (in the form of Ito)\footnote{%
Following the standard notation of the SDE theory, we use capital letters to
denote the SDE solution and small letters for the initial data and for
corresponding \textquotedblleft dummy\textquotedblright\ variables.}
\begin{eqnarray}
dR^{j} &=&\frac{P^{j}}{m}dt,\ \ R^{j}(0)=r^{j},  \label{a4} \\
dP^{j} &=&-\nabla _{r^{j}}U(\mathbf{R},\mathbf{Q})dt  \notag \\
&&-\gamma g(P^{j},R^{j})dt+b(R^{j})dw^{j}(t),\ \ P^{j}(0)=p^{j},  \notag \\
dQ^{j} &=&\sum_{l=1}^{3}\nabla _{\pi ^{j}}V_{l}(Q^{j},\Pi ^{j})dt,\ \
Q^{j}(0)=q^{j},\ \ |q^{j}|=1,  \notag \\
d\Pi ^{j} &=&-\sum_{l=1}^{3}\nabla _{q^{j}}V_{l}(Q^{j},\Pi ^{j})dt-\nabla
_{q^{j}}U(\mathbf{R},\mathbf{Q})dt  \label{a5} \\
&-&\!\!\Gamma G(Q^{j},\Pi ^{j})dt+B(Q^{j},\Pi ^{j})dW^{j}(t),\ \ \Pi
^{j}(0)=\pi ^{j},  \notag \\
j &=&1,\ldots ,n,  \notag
\end{eqnarray}%
where $\gamma \geq 0$ and $\Gamma \geq 0$ with $\gamma \Gamma >0$ 
are the friction coefficients for the translational and rotational motions,
respectively, measured in units of inverse time, which control the strength
of coupling of the system to the \textquotedblleft heat
bath\textquotedblright ; $g$ is a $3$-dimensional appropriately normalized
vector; $G$ is a $4$-dimensional vector, which provides a balance in
coupling various rotational degrees of freedom with the \textquotedblleft
heat bath\textquotedblright ; $b$ and $B$ are $3$-by-$3$ and $4$-by-$4$
matrices, respectively; and $(\mathbf{w}^{\mathsf{T}},\mathbf{W}^{\mathsf{T}%
})^{\mathsf{T}}=(w^{1\,\mathsf{T}},\ldots ,w^{n\,\mathsf{T}},W^{1^{^{%
\mathsf{T}}}},\ldots ,W^{n\,\mathsf{T}})^{\mathsf{T}}$ is a $(3n+4n)$%
-dimensional standard Wiener process with $%
w^{j}=(w_{1}^{j},w_{2}^{j},w_{3}^{j})^{\mathsf{T}}$ and $%
W^{j}=(W_{0}^{j},W_{1}^{j},W_{2}^{j},W_{3}^{j})^{\mathsf{T}}.$

For simplicity, we assumed here that $g$, $G$, $b$, and $B$ are the same for
all $n$ molecules, although one could choose them depending on the molecule
number $j$. The latter can be especially useful for systems consisting of
significantly different types of molecules. It is also natural to require
that each degree of freedom is thermalized by its own independent noise, and
in what follows we assume that the matrices $b$ and $B$ are diagonal.
Further, we suppose that the coefficients of (\ref{a4})-(\ref{a5}) are
sufficiently smooth functions and the process $X(t)=(\mathbf{R}^{\mathsf{T}%
}(t),\mathbf{P}^{\mathsf{T}}(t),\mathbf{Q}^{\mathsf{T}}(t),\mathbf{\Pi }^{%
\mathsf{T}}(t))^{\mathsf{T}}$ is ergodic, i.e., there exists a unique
invariant measure $\mu $ of $X$ and independently of $x\in \mathbb{R}^{14n}$
there exists the limit%
\begin{equation}
\lim_{t\rightarrow \infty }E\varphi (X(t;x))=\int \varphi (x)\,d\mu
(x):=\varphi ^{erg}  \label{PA31}
\end{equation}%
for any function $\varphi (x)$ with polynomial growth at infinity (see Refs.~%
\onlinecite{Has,Soize,Stua,Tal02} and references therein). Here $X(t;x)$ is
the solution $X(t)$ of (\ref{a4})-(\ref{a5}) with the initial condition $%
X(0)=X(0;x)=x.$

It is not difficult to see that the solution of (\ref{a4})-(\ref{a5})
preserves the property (\ref{a21}), i.e.,
\begin{equation}
|Q^{j}(t)|=1,\ \ j=1,\ldots ,n\,,\ \ \mbox{for all~}t\geq 0.  \label{a211}
\end{equation}

Now we find relations between $\gamma $, $\Gamma $, $g$, $G$, $b$, and $B$
such that the invariant measure $\mu $ is Gibbsian with the density (\ref{a3}%
). The density $\rho (\mathbf{r},\mathbf{p},\mathbf{q},\bm{\pi})$ should
satisfy the stationary Fokker-Planck equation:
\begin{equation}
L^{\ast }\rho =0,  \label{a6}
\end{equation}%
where
\begin{widetext}
\begin{equation*}\begin{split}
L^{\ast }\rho &:=\sum_{j=1}^{n}\Biggl\{ \sum_{i=1}^{3}%
\frac{b_{ii}^{2}(r^{j})}{2}\frac{\partial ^{2}\rho }{\bigl( \partial p_{i}^{j}\bigr) ^{2}}%
+\sum_{i=1}^{4}\frac{1}{2}\frac{\partial ^{2}}{\bigl( \partial \pi_{i}^{j}\bigr)^{2}}
\left( B_{ii}^{2}(q^{j},\pi ^{j})\rho \right)
 +\nabla _{p^{j}}\cdot\left[ \gamma g(p^{j},r^{j}\mathbf{)}\rho \right]%
-\nabla _{q^{j}}\cdot \left( \nabla_{\pi^{j}}\sum_{l=1}^{3}V_{l}(q^{j},\pi ^{j})\rho \right)\\
&\left.%
+\nabla _{\pi ^{j}}\cdot \left[ \left( %
\nabla _{q^{j}}\sum_{l=1}^{3}V_{l}(q^{j},\pi ^{j})%
+\nabla _{q^{j}}U(\mathbf{r}, \mathbf{q})%
+\Gamma  G(q^{j},\pi ^{j})\right) \rho \right] \right\}%
-\frac{1}{m}\nabla _{\mathbf{r}}\cdot (\mathbf{p}\rho )%
+\nabla_{\mathbf{p}}\cdot \left[ \nabla _{\mathbf{r}}U(\mathbf{r},\mathbf{q})\rho \right].
\end{split}\end{equation*}
\end{widetext}After some calculations, we get the required relations:
\begin{equation}
\frac{\beta }{m}\frac{b_{ii}^{2}}{2}\left[ \frac{\beta }{m}\bigl(p_{i}^{j}%
\bigr)^{2}-1\right] +\gamma \frac{\partial g_{i}}{\partial p_{i}^{j}}-\gamma
\frac{\beta }{m}g_{i}p_{i}^{j}=0  \label{a7}
\end{equation}%
and
\begin{equation}
\begin{split}
\beta \frac{B_{ii}^{2}}{2}& \Biggl[\beta \biggl(\frac{\partial H}{\partial
\pi _{i}^{j}}\biggr)^{2}-\frac{\partial ^{2}H}{\bigl(\partial \pi _{i}^{j}%
\bigl)^{2}}\Biggr]+\biggl(\frac{\partial B_{ii}}{\partial \pi _{i}^{j}}%
\biggr)^{2} \\
& +B_{ii}\frac{\partial ^{2}B_{ii}}{\bigl(\partial \pi _{i}^{j}\bigr)^{2}}%
-\beta B_{ii}\frac{\partial B_{ii}}{\partial \pi _{i}^{j}}\frac{\partial H}{%
\partial \pi _{i}^{j}} \\
& +\Gamma \frac{\partial G_{i}}{\partial \pi _{i}^{j}}-\Gamma \beta G_{i}%
\frac{\partial H}{\partial \pi _{i}^{j}}=0.
\end{split}
\label{a8}
\end{equation}

Since numerical methods are usually simpler for systems with additive noise,
we limit computational consideration of thermostats in this paper to the
case of $b_{ii}$ and $B_{ii}$ both being constant. At the same time, we note
that general thermostats (\ref{a4})-(\ref{a5}) with (\ref{a7})-(\ref{a8})
may have some beneficial features for certain systems but we leave this
question for further study. For constant $b_{ii}$ and $B_{ii},$ the
relations (\ref{a7})-(\ref{a8}) take the form
\begin{equation*}
\gamma g_{i}(p^{j},r^{j}\mathbf{)}=\frac{\beta }{m}\frac{b_{ii}^{2}}{2}%
p_{i}^{j}\ \ \text{and\ \ }\Gamma G_{i}(q^{j},\pi ^{j})=\beta \frac{%
B_{ii}^{2}}{2}\frac{\partial H}{\partial \pi _{i}^{j}}.
\end{equation*}%
In the considered molecular model it is natural to have the same value for
all $b_{ii}$, $i=1,2,3,$ and the same value for all $B_{ii}$, $i=1,\ldots
,4. $ Further, taking into account the form of the Hamiltonian (\ref{a1})
and that
\begin{eqnarray*}
\nabla _{\pi }\sum_{l=1}^{3}V_{l}(q,\pi )&=&\frac{1}{4}\sum_{l=1}^{3} \frac{1%
}{I_{l}}\left[ \pi ^{\mathsf{T}}S_{l}q\right] S_{l}q \\
&=&\frac{1}{4}\sum_{l=1}^{3}\frac{1}{I_{l}}S_{l}q\left[ S_{l}q\right]^{%
\mathsf{T}}\pi ,
\end{eqnarray*}%
we can write
\begin{equation*}
G(q,\pi )=J(q)\pi \ \ \text{and}\ \ B_{ii}^{2}=\frac{2M\Gamma }{\beta }
\end{equation*}%
with
\begin{equation}  \label{a15}
J(q)=\frac{\sum_{l=1}^{3}\frac{1}{I_{l}}S_{l}q\left[ S_{l}q\right] ^{\mathsf{%
T}}}{\sum_{l=1}^{3}\frac{1}{I_{l}}}\ \ \text{and\ \ } M=\frac{4}{%
\sum_{l=1}^{3}\frac{1}{I_{l}}}.
\end{equation}%
Note that $\limfunc{Tr}J(q)=|q|^{2}=1.$

Thus, in this paper under `\textit{Langevin thermostat}' we understand the
following stochastic system
\begin{eqnarray}
dR^{j} &=&\frac{P^{j}}{m}dt,\ \ R^{j}(0)=r^{j},  \label{lt1} \\
dP^{j} &=&-\nabla _{r^{j}}U(\mathbf{R},\mathbf{Q})dt  \notag \\
&&-\gamma P^{j}dt+\sqrt{\frac{2m\gamma }{\beta }}dw^{j}(t),\ \
P^{j}(0)=p^{j},  \notag \\
dQ^{j} &=&\sum_{l=1}^{3}\nabla _{\pi ^{j}}V_{l}(Q^{j},\Pi ^{j})dt,\
Q^{j}(0)=q^{j},\ |q^{j}|=1,  \label{lt2} \\
d\Pi ^{j} &=&-\sum_{l=1}^{3}\nabla _{q^{j}}V_{l}(Q^{j},\Pi ^{j})dt-\nabla
_{q^{j}}U(\mathbf{R},\mathbf{Q})dt  \notag \\
&&-\Gamma J(Q^{j})\Pi ^{j}dt+\sqrt{\frac{2M\Gamma }{\beta }}dW^{j}(t),\ \
\Pi ^{j}(0)=\pi ^{j},  \notag \\
j &=&1,\ldots ,n,  \notag
\end{eqnarray}%
where $J(q)$ and $M$ are from (\ref{a15}), the rest of the notation is as in
(\ref{a4})-(\ref{a5}). We recall that $\gamma $ and $\Gamma $ are free
parameters having the physical meaning of the 
strength of coupling to the heat bath. 

Let us fix a molecule and write the equations for the body-fixed angular
velocities $\omega _{x}$, $\omega _{y}$, and $\omega _{z}$ corresponding to
the rotational Langevin subsystem (\ref{lt2}). To this end, we recall\cite%
{qua02} that
\begin{equation*}
\omega _{x}=2(S_{1}Q)^{\mathsf{T}}\dot{Q},\ \ \omega _{y}=2(S_{2}Q)^{\mathsf{%
T}}\dot{Q},\ \ \omega _{z}=2(S_{3}Q)^{\mathsf{T}}\dot{Q}.
\end{equation*}%
Then we obtain%
\begin{equation}\label{av}\begin{split}
d\omega _{x} &=\left( \frac{\tau _{1}}{I_{1}}+\frac{I_{2}-I_{3}}{I_{1}}%
\omega _{y}\omega _{z}\right) dt-\frac{M\Gamma }{4I_{1}}\omega _{x}dt\\
&+\frac{1}{I_{1}}\sqrt{\frac{2M\Gamma }{\beta }}d\Upsilon _{1}, \\
d\omega _{y} &=\left( \frac{\tau _{2}}{I_{2}}+\frac{I_{3}-I_{1}}{I_{2}}%
\omega _{x}\omega _{z}\right) dt-\frac{M\Gamma }{4I_{2}}\omega _{y}dt \\
&+\frac{1}{I_{2}}\sqrt{\frac{2M\Gamma }{\beta }}d\Upsilon _{2},\\
d\omega _{z} &=\left( \frac{\tau _{3}}{I_{3}}+\frac{I_{1}-I_{2}}{I_{3}}%
\omega _{x}\omega _{y}\right) dt-\frac{M\Gamma }{4I_{3}}\omega _{z}dt \\
&+\frac{1}{I_{3}}\sqrt{\frac{2M\Gamma }{\beta }}d\Upsilon _{3},
\end{split}\end{equation}%
where $\tau _{i}$ are the torques, $\tau _{i}=-\frac{1}{2}(S_{i}Q)^{\mathsf{T%
}}\nabla _{q}U,$ and $d\Upsilon _{i}=\frac{1}{2}\sum_{j=1}^{4}(S_{i}Q)_{j}\
dW_{j},$ which can be interpreted as random torques. For $\Gamma =0,$ (\ref%
{av}) coincide with the equations for the angular velocities in Ref.~%
\onlinecite{qua02}. We also note that due to the form of the Hamiltonian (%
\ref{a1}) the auxiliary velocity $\omega_{0}$ used in Ref.~\onlinecite{qua02}
in the derivation of the Hamiltonian system for rigid-body dynamics is
identically equal to zero.

\subsection{A mixture of gradient system and Langevin-type equation\label%
{sec_grad}}

Another possibility of stochastic thermostating of (\ref{a20}) rests on a
mixture of a gradient system for the translational dynamics and
Langevin-type equation for the rotational dynamics. We note that according
to the density of Gibbsian measure (\ref{a3}) the center-of-mass momenta $%
\mathbf{P}$ are independent Gaussian random variables and they are
independent of the other components of the system, so we can avoid
simulating $\mathbf{P}$ via a differential equation.

Consider the `\textit{gradient-Langevin thermostat}'
\begin{align}
d\mathbf{R}& =-\frac{\nu }{m}\nabla _{\mathbf{r}}U(\mathbf{R},\mathbf{Q})dt+%
\sqrt{\frac{2\nu }{m\beta }}d\mathbf{w}(t),\ \ \mathbf{R}(0)=\mathbf{r},
\label{a10} \\
dQ^{j}& =\nabla _{\pi ^{j}}\sum_{l=1}^{3}V_{l}(Q^{j},\Pi ^{j})dt,\ \
Q^{j}(0)=q^{j},\ \ |q^{j}|=1,  \label{a100} \\
d\Pi ^{j}& =-\nabla _{q^{j}}\sum_{l=1}^{3}V_{l}(Q^{j},\Pi ^{j})dt-\nabla
_{q^{j}}U(\mathbf{R},\mathbf{Q})dt  \notag \\
& -\Gamma J(Q^{j})\Pi ^{j}dt+\sqrt{\frac{2M\Gamma }{\beta }}dW^{j}(t),\ \
\Pi ^{j}(0)=\pi ^{j},  \notag \\
j& =1,\ldots ,n,  \notag
\end{align}%
where all the notation is as in (\ref{lt1})-(\ref{lt2}) and, in particular, $%
J(q)$\ and $M$ are from (\ref{a15}). The invariant measure of (\ref{a10})-(%
\ref{a100}) is (\ref{a3}) integrated over $\mathbf{p}.$ The property (\ref%
{a211}) is preserved. The gradient-Langevin thermostat has two free
parameters, $\nu > 0$ and $\Gamma \geq 0$. The latter is the same as in (\ref%
{lt1})-(\ref{lt2}), while the former, measured in units of time, controls
the speed of evolution of the gradient subsystem (\ref{a10}).

It is important to note that the gradient system does not have a natural
dynamical time evolution similar to Hamiltonian or Langevin dynamics. This
is because changing parameter $\nu $ simply leads to a time renormalization
of the gradient subsystem (\ref{a10}). However, when linked with the
Langevin dynamics for rotational degrees of freedom, as in (\ref{a10})-(\ref%
{a100}), parameter $\nu $ controls the \textquotedblleft
speed\textquotedblright\ of evolution of the gradient subsystem relative to
the speed of the rotational dynamics.


To check that
\begin{equation*}
\rho (\mathbf{r},\mathbf{q},\bm{\pi})\varpropto \exp \Bigl(-\beta \Bigl[%
\sum_{j=1}^{n}\sum_{l=1}^{3}V_{l}(q^{j},\pi ^{j})+U(\mathbf{r},\mathbf{q})%
\Bigr]\Bigr)
\end{equation*}%
is the density of the invariant measure for (\ref{a10})-(\ref{a100}), one
needs to consider the Fokker-Planck equation (\ref{a6}) with the following
operator:
\begin{widetext}
\begin{equation*}\begin{split}
L^{\ast }\rho &:=\sum_{j=1}^{n}\Biggl\{ \frac{\nu}{m \beta } \sum_{i=1}^{3}%
\frac{\partial ^{2}\rho }{\bigl( \partial r_{i}^{j}\bigr)^{2}}+\frac{%
M\Gamma }{\beta }\sum_{i=1}^{4}\frac{\partial ^{2}}{\bigl( \partial \pi
_{i}^{j}\bigr)^{2}}\rho +\nabla _{\pi ^{j}}\cdot \left[ \left(
\nabla _{q^{j}}\sum_{l=1}^{3}
V_{l}(q^{j},\pi ^{j})+\nabla _{q^{j}}U(\mathbf{r},\mathbf{q})+
\Gamma J(q^{j})\pi^{j}\right) \rho \right] \\
&\left. -\nabla _{q^{j}}\cdot \left( \nabla _{\pi
^{j}}\sum_{l=1}^{3}V_{l}(q^{j},\pi ^{j})\rho \right) \right\} +\frac{\nu}{m} \nabla
_{\mathbf{r}}\cdot \left[ \nabla _{\mathbf{r}}U(\mathbf{r},\mathbf{q})\rho %
\right] .
\end{split}\end{equation*}
\end{widetext}

Let us remark\cite{Nel67} that the gradient sub-system (\ref{a10}) can be
viewed as an overdamped limit of the Langevin translational sub-system (\ref%
{lt1}) for a fixed $\mathbf{Q}$.

\section{Numerical integrators\label{sec_num}}

In this section we consider effective second-order numerical methods for the
Langevin thermostat (\ref{lt1})-(\ref{lt2}) and the gradient-Langevin
thermostat (\ref{a10})-(\ref{a100}). We first recall the idea of
quasi-symplectic integrators for Langevin-type equations introduced in Ref.~%
\onlinecite{MT5} (see also Refs.~\onlinecite{MT6,MT7}) and also some basic
facts from stochastic numerics\cite{MT6}.

Consider the Langevin equations (\ref{lt1})-(\ref{lt2}). Let $D_{0}\in
\mathbb{R}^{d},$ $d=14n,$ be a domain with finite volume. The transformation
$x=(\mathbf{r}^{\mathsf{T}},\mathbf{p}^{\mathsf{T}},\mathbf{q}^{\mathsf{T}},%
\mathbf{\pi }^{\mathsf{T}})^{\mathsf{T}}\mapsto X(t)=X(t;x)=(\mathbf{R}^{%
\mathsf{T}}(t;x),\mathbf{P}^{\mathsf{T}}(t;x),\mathbf{Q}^{\mathsf{T}}(t;x),%
\mathbf{\Pi }^{\mathsf{T}}(t;x))^{\mathsf{T}}$ maps $D_{0}$ into the domain $%
D_{t}.$ The volume $V_{t}$ of the domain $D_{t}$ is equal to
\begin{eqnarray}
V_{t} &=&\int\limits_{D_{t}}dX^{1}\ldots dX^{d}  \label{vol0} \\
&=&\int\limits_{D_{0}}\left\vert \frac{D(X^{1},\ldots ,X^{d})}{%
D(x^{1},\ldots ,x^{d})}\right\vert \,dx^{1}\ldots dx^{d}.  \notag
\end{eqnarray}%
The Jacobian determinant $\mathbb{J}$ is equal to (see, e.g., Ref.~%
\onlinecite{hadd}):
\begin{equation}
\mathbb{J}=\frac{D(X^{1},\ldots ,X^{d})}{D(x^{1},\ldots ,x^{d})}=\exp \left(
-n(3\gamma +\Gamma )\cdot t\right) .  \label{jac1}
\end{equation}%
The system (\ref{lt1})-(\ref{lt2}) preserves phase volume when $\gamma =0$
and $\Gamma =0$. If $\gamma \geq 0$ and $\Gamma \geq 0$ with $\gamma \Gamma >0$ 
then phase-volume contractivity takes place.

If we omit the damping terms, $-\gamma P^{j}$ and $-\Gamma J\Pi ^{j},$ in (%
\ref{lt1})-(\ref{lt2}) then the system becomes a Hamiltonian system with
additive noise\cite{hadd,MT6}, i.e., its phase flow preserves symplectic
structure. Under $\gamma =0$ and $\Gamma =0$, (\ref{lt1})-(\ref{lt2}) takes
the form of the deterministic Hamiltonian system (\ref{a20}).

We say that the method based on a one-step approximation $\bar{X}=\bar{X}%
(t+h;t,x)$, $h>0$, is symplectic if $\bar{X}$ preserves symplectic structure%
\cite{hadd,MT6}. It is natural to expect that making use of numerical
methods, which are close, in a sense, to symplectic ones, has advantages
when applying to stochastic systems close to Hamiltonian ones. In Ref.~%
\onlinecite{MT5} (see also Ref.~\onlinecite{MT6}) numerical methods (they
are called quasi-symplectic) for Langevin equations were proposed, which
satisfy the two structural conditions:

\begin{enumerate}
\item[\textbf{RL1.}] \textit{The method applied to Langevin equations
degenerates to a symplectic method when the Langevin system degenerates to a
Hamiltonian one.}

\item[\textbf{RL2.}] \textit{The Jacobian determinant }$\mathbb{\bar{J}}=D%
\bar{X}/Dx$ \textit{does not depend on} $x.$
\end{enumerate}

The requirement RL1 ensures closeness of quasi-symplectic integrators to the
symplectic ones. As it is always assumed, a method is convergent and,
consequently, $\mathbb{\bar{J}}$ is close to $\mathbb{J}$ at any rate. The
requirement RL2 is natural since the Jacobian $\mathbb{J}$ of the original
system (\ref{lt1})-(\ref{lt2}) does not depend on $x.$ RL2 reflects the
structural properties of the system which are connected with the law of
phase volume contractivity. It is often possible to reach a stronger
property consisting in the equality $\mathbb{\bar{J}}=\mathbb{J}.$

We usually consider two types of numerical methods for SDEs: mean-square and
weak\cite{MT6}. Mean-square methods are useful for direct simulation of
stochastic trajectories while weak methods are sufficient for evaluation of
averages and are simpler than mean-square ones. Therefore, weak methods are
most suitable for the purposes of this paper. Let us recall\cite{MT6} that a
method $\bar{X}$ is weakly convergent with order $p>0$ if
\begin{equation}
|E\varphi (\bar{X}(T))-E\varphi (X(T))|\leq Ch^{p},  \label{A20}
\end{equation}%
where $h>0$ is a time discretization step and $\varphi $ is a sufficiently
smooth function with growth at infinity not faster than polynomial. The
constant $C$ does not depend on $h,$ it depends on the coefficients of a
simulated stochastic system, on $\varphi ,$ and $T.$

\subsection{Numerical schemes for the Langevin thermostat\label{sec_numla}}

We assume that the system (\ref{lt1})-(\ref{lt2}) has to be solved on a time
interval $[0,T]$ and for simplicity we use a uniform time discretization
with the step $h=T/N.$ Using standard ideas of stochastic numerics\cite%
{MT5,MT6} including splitting techniques and the numerical method from Ref.~%
\onlinecite{qua02} for the deterministic Hamiltonian system (\ref{a20}), we
derive two quasi-symplectic integrators for the Langevin system (\ref{lt1})-(%
\ref{lt2}).

The first integrator (Langevin A) is based on splitting the Langevin system (%
\ref{lt1})-(\ref{lt2}) into the Hamiltonian system with additive noise
(i.e., (\ref{lt1})-(\ref{lt2}) without the damping terms) and the
deterministic system of linear differential equations of the form%
\begin{equation}
\begin{split}
\dot{\mathbf{p}}& =-\gamma \mathbf{p} \\
\dot{\pi}^{j}& =-\Gamma J(q^{j})\pi ^{j},\ j=1,\ldots ,n\,.
\end{split}
\label{nsl1}
\end{equation}%
We construct a second-order weak quasi-symplectic integrator for the
stochastic Hamiltonian system\cite{hadd,MT6} and appropriately concatenate%
\cite{MT5,MT6} it with the exact solution of (\ref{nsl1}). The resulting
numerical method is given below.

Introduce the mapping $\Psi _{l}(t;q,\pi ):$ $(q,\pi )\mapsto (\mathcal{Q},%
\mathit{\Pi })$ defined by 
\begin{equation}
\begin{split}
\mathcal{Q}& =\cos (\chi _{l}t)q+\sin (\chi _{l}t)S_{l}q\,, \\
\mathit{\Pi }& =\cos (\chi _{l}t)\pi +\sin (\chi _{l}t)S_{l}\pi \,,
\end{split}
\label{a24}
\end{equation}%
where
\begin{equation*}
\chi _{l}=\frac{1}{4I_{l}}\pi ^{\mathsf{T}}S_{l}q\,.
\end{equation*}%
The first quasi-symplectic scheme for (\ref{lt1})-(\ref{lt2}) can be written
in the form:

\begin{center}\underline{Langevin A}\end{center}
\begin{eqnarray}
\mathbf{P}_{0} &=&\mathbf{p},\ \ \mathbf{R}_{0}=\mathbf{r},\ \mathbf{Q}_{0}=%
\mathbf{q},\ \ \mathbf{\Pi }_{0}=\mathbf{\pi ,}  \label{firla} \\
\mathcal{P}_{1,k} &=&\mathbf{P}_{k}\exp ( -\gamma h/2) \,,  \notag
\\
\mathit{\Pi }_{1,k}^{j} &=&\exp \big( -\Gamma J(Q_{k}^{j})h/2\big) \Pi
_{k}^{j},\ \ j=1,\ldots ,n,  \notag
\end{eqnarray}%
\begin{eqnarray*}
\mathcal{P}_{2,k} &=&\mathcal{P}_{1,k}-\frac{h}{2}\nabla _{\mathbf{r}}U(%
\mathbf{R}_{k},\mathbf{Q}_{k})+\frac{\sqrt{h}}{2}\sqrt{\frac{2m\gamma }{%
\beta }}\mathbf{\xi }_{k} \\
\mathit{\Pi }_{2,k}^{j} &=&\mathit{\Pi }_{1,k}^{j}-\frac{h}{2}\nabla
_{q^{j}}U(\mathbf{R}_{k},\mathbf{Q}_{k})+\frac{\sqrt{h}}{2}\sqrt{\frac{%
2M\Gamma }{\beta }}\eta _{k}^{j} \\
&&-\frac{h^{2}}{4}\frac{\Gamma }{\beta }Q_{k}^{j},\ \ j=1,\ldots ,n, \\
\mathbf{R}_{k+1} &=&\mathbf{R}_{k}+\frac{h}{m}\mathcal{P}_{2,k},
\end{eqnarray*}%
\begin{eqnarray*}
(\mathcal{Q}_{1,k}^{j},\mathit{\Pi }_{3,k}^{j}) &=&\Psi _{3}(h/2;Q_{k}^{j},%
\mathit{\Pi }_{2,k}^{j}), \\
(\mathcal{Q}_{2,k}^{j},\mathit{\Pi }_{4,k}^{j}) &=&\Psi _{2}(h/2;\mathcal{Q}%
_{1,k}^{j},\mathit{\Pi }_{3,k}^{j}), \\
(\mathcal{Q}_{3,k}^{j},\mathit{\Pi }_{5,k}^{j}) &=&\Psi _{1}(h;\mathcal{Q}%
_{2,k}^{j},\mathit{\Pi }_{4,k}^{j}), \\
(\mathcal{Q}_{4,k}^{j},\mathit{\Pi }_{6,k}^{j}) &=&\Psi _{2}(h/2;\mathcal{Q}%
_{3,k}^{j},\mathit{\Pi }_{5,k}^{j}), \\
(Q_{k+1}^{j},\mathit{\Pi }_{7,k}^{j}) &=&\Psi _{3}(h/2;\mathcal{Q}_{4,k}^{j},%
\mathit{\Pi }_{6,k}^{j}),\ \ j=1,\ldots ,n,
\end{eqnarray*}%
\begin{eqnarray}
\mathit{\Pi }_{8,k}^{j} &=&\mathit{\Pi }_{7,k}^{j}-\frac{h}{2}\nabla
_{q^{j}}U(\mathbf{R}_{k+1},\mathbf{Q}_{k+1})  \notag \\
&&+\frac{\sqrt{h}}{2}\sqrt{\frac{2M\Gamma }{\beta }}\eta _{k}^{j}-\frac{h^{2}%
}{4}\frac{\Gamma }{\beta }Q_{k+1}^{j},\ j=1,\ldots ,n,  \notag \\
\mathcal{P}_{3,k} &=&\mathcal{P}_{2,k}-\frac{h}{2}\nabla _{\mathbf{r}}U(%
\mathbf{R}_{k+1},\mathbf{Q}_{k+1})+\frac{\sqrt{h}}{2}\sqrt{\frac{2m\gamma }{%
\beta }}\mathbf{\xi }_{k},  \notag
\end{eqnarray}%
\begin{eqnarray*}
\mathbf{P}_{k+1} &=&\mathcal{P}_{3,k}\exp ( -\gamma h/2) , \\
\Pi _{k+1}^{j} &=&\exp \big( -\Gamma J(Q_{k+1}^{j})h/2\big) \mathit{\Pi }%
_{8,k}^{j},\ \ j=1,\ldots ,n, \\
k &=&0,\ldots ,N-1,
\end{eqnarray*}%
where $\mathbf{\xi }_{k}=(\xi _{1,k},\ldots ,\xi _{3n,k})^{\mathsf{T}}$ and $%
\eta _{k}^{j}=(\eta _{1,k}^{j},\ldots ,\eta _{4,k}^{j})^{\mathsf{T}},$ $%
j=1,\ldots ,n,$ with their components being i.i.d. with the same law
\begin{equation}
P(\theta =0)=2/3,\ \ P(\theta =\pm \sqrt{3})=1/6.  \label{n31}
\end{equation}

It is easy to check\cite{MT5,MT6} that the scheme (\ref{firla}) is
quasi-symplectic. Moreover, the Jacobian $\mathbb{\bar{J}}$ of the
corresponding one-step approximation is exactly equal to the Jacobian $%
\mathbb{J}$ of the original system (\ref{lt1})-(\ref{lt2}).

To prove the second order of weak convergence of (\ref{firla})-(\ref{n31}),
we compared the corresponding one-step approximation with the one-step
approximation corresponding to the standard second-order weak method for
SDEs with additive noise from Ref.~\onlinecite{MT6}[p. 113]. The following
properties are used in this proof:%
\begin{equation*}
2M\sum_{i=0}^{3}\frac{\partial ^{2}}{\partial \pi _{i}^{2}}%
\sum_{l=1}^{3}\nabla _{q}V_{l}(q,\pi )=4q,
\end{equation*}%
\begin{equation*}
\frac{\partial ^{2}}{\partial \pi _{i}\partial \pi _{j}}\nabla _{\pi
}V_{l}(q,\pi )=0,
\end{equation*}%
and
\begin{equation*}
\frac{\partial }{\partial \pi _{i}^{l}}V_{l}(q^{j},\pi ^{j})=\frac{\partial
}{\partial q_{i}^{l}}V_{l}(q^{j},\pi ^{j})=0\ \ \text{for }j\neq l.
\end{equation*}%
As it is usual in stochastic numerics\cite{MT6}, we prove convergence of a
numerical method under the global Lipschitz assumption on the coefficients
of the stochastic system, which can then be relaxed using the concept of
rejecting exploding trajectories\cite{MT7}.

Analogously to the deterministic case\cite{qua02}, one can verify that the
scheme (\ref{firla}) preserves (\ref{a211}), i.e., $|Q_{k}^{j}|=1,\ \
j=1,\ldots ,n\,,$ for all $k.$ We summarize the properties of the method (%
\ref{firla})-(\ref{n31}) in the following statement.

\begin{proposition}
\label{prp1}The numerical scheme $(\ref{firla})$-$(\ref{n31})$ for $(\ref%
{lt1})$-$(\ref{lt2})$ is quasi-symplectic, it preserves the structural
property $(\ref{a211})$, and is of weak order two.
\end{proposition}

We note that one can choose $\mathbf{\xi }_{k}=(\xi _{1,k},\ldots ,\xi
_{3n,k})^{\mathsf{T}}$ and $\eta _{k}^{j}=(\eta _{1,k}^{j},\ldots ,\eta
_{4,k}^{j})^{\mathsf{T}},$ $j=1,\ldots ,n,$ so that their components are
i.i.d. Gaussian random variables with zero mean and unit variance. In this
case the weak order of the scheme remains second as when we use the simple
discrete distribution (\ref{n31}). Since simulation of the discrete random
variables is cheaper than Gaussian ones, it is preferable to use (\ref{n31})
and it was used in all our experiments in this paper. Let us remark in
passing that in the case of Gaussian random variables the above scheme also
converges in the mean-square sense\cite{MT6} with order one.

Note that $\exp (-\Gamma J(q)h/2)$ in (\ref{firla}) is the exponent of a
matrix. It can be computed using a standard linear algebra package (such as
LAPACK). Since $J(q)$ is a symmetric matrix, LAPACK's \texttt{dsyev} routine
can be used to obtain the eigen decomposition
\begin{equation}
J(q)=T(q)\Lambda _{J}(q)T^{\mathsf{T}}(q)\,,  \label{eq:eigJq}
\end{equation}%
where $T(q)$ is a matrix whose columns are the eigenvectors of $J(q)$ and
\begin{equation*}
\Lambda _{J}(q)=\mathrm{diag}(\lambda _{J,1},\ldots ,\lambda _{J,4})
\end{equation*}%
is a diagonal matrix of the corresponding eigenvalues. Then
\begin{equation*}
\exp (-\Gamma J(q)h/2)=T(q)\exp (-\Gamma \Lambda _{J}(q)h/2)T^{\mathsf{T}%
}(q)\,,
\end{equation*}%
where
\begin{equation*}
\exp (-\Gamma \Lambda _{J}(q)h/2)=\mathrm{diag}\bigl(\mathrm{e}^{-\Gamma
\lambda _{J,1}h/2},\ldots ,\mathrm{e}^{-\Gamma \lambda _{J,4}h/2}\bigr).
\end{equation*}%
Alternatively, the matrix exponent $\exp (-\Gamma J(q)h/2)$ in (\ref{firla})
can be approximated via the Taylor expansion. To ensure the second-order
convergence, it is sufficient to approximate it with accuracy $O(h^{3})$ at
one step; the scheme will remain quasi-symplectic but the Jacobian $\mathbb{%
\bar{J}}$ will no longer be equal to $\mathbb{J}$ in (\ref{jac1}).

When the parameters $\gamma $ and $\Gamma $ are large (the strong coupling
to the \textquotedblleft heat bath\textquotedblright\ conditions), we
propose to use a numerical integrator for the Langevin system
(\ref{lt1})-(\ref{lt2}) based on the following splitting:
\begin{equation}
\begin{split}
d\mathbf{P}_{I}& =-\gamma \mathbf{P}_{I}\,dt+\sqrt{\frac{2m\gamma }{\beta }}d%
\mathbf{w}(t), \\
d\Pi _{I}^{j}& =-\Gamma J(q)\Pi _{I}^{j}dt+\sqrt{\frac{2M\Gamma }{\beta }}%
dW^{j}(t);
\end{split}
\label{lb1}
\end{equation}%
\begin{equation}\label{lb2} \begin{split}
d\mathbf{R}_{II} =& \frac{\mathbf{P}_{II}}{m}\,dt   \\
d\mathbf{P}_{II} =& -\nabla _{\mathbf{r}}U(\mathbf{R}_{II},\mathbf{Q}_{II})dt, \\
dQ_{II}^{j} =& \nabla _{\pi ^{j}}\sum_{l=1}^{3}V_{l}(Q_{II}^{j},\Pi
_{II}^{j})dt\,, \\
d\Pi _{II}^{j} =& -\nabla _{q^{j}}U(\mathbf{R}_{II},\mathbf{Q}_{II})dt\\
 &-\nabla _{q^{j}}\sum_{l=1}^{3}V_{l}(Q_{II}^{j},\Pi _{II}^{j})dt\,, \\
j =& 1,\ldots ,n.
\end{split}\end{equation}%
The SDEs (\ref{lb1}) have the exact solution:
\begin{equation}
\begin{split}
\mathbf{P}_{I}(t)& =\mathbf{P}_{I}(0)e^{-\gamma t}+\sqrt{\frac{2m\gamma }{%
\beta }}\int_{0}^{t}e^{-\gamma (t-s)}d\mathbf{w}(s), \\
\Pi _{I}^{j}(t)& =\exp (-\Gamma J(q)t)\Pi _{I}^{j}(0) \\
& +\sqrt{\frac{2M\Gamma }{\beta }}\int_{0}^{t}\exp (-\Gamma
J(q)(t-s))dW^{j}(s).
\end{split}
\label{lbe}
\end{equation}%
To construct a method based on the splitting (\ref{lb1})-(\ref{lb2}), we
take half a step of (\ref{lb1}) using (\ref{lbe}), one step of a symplectic
method for (\ref{lb2}), and again half a step of (\ref{lb1}).

The Ito integral in the expression for $\Pi _{I}^{j}$ in (\ref{lbe}) is a
four-dimensional Gaussian vector with zero mean and the covariance matrix
\begin{equation}
\begin{split}
C(t;q)& =\frac{2M\Gamma }{\beta }\int_{0}^{t}\exp [-2\Gamma J(q)(t-s)]ds \\
& =\frac{M}{\beta }T(q)\Lambda _{C}(t;q,\Gamma )T^{\mathsf{T}}(q),
\end{split}
\label{eq:Crot}
\end{equation}%
where $T(q)$ is as in (\ref{eq:eigJq}) and
\begin{equation*}
\Lambda _{C}(t;q,\Gamma )=\mathrm{diag}(\lambda _{C,1},\ldots ,\lambda
_{C,4})
\end{equation*}%
with
\begin{equation}
\begin{split}
\lambda _{C,i}(t;q,\Gamma )& =\left\{
\begin{array}{cc}
2\Gamma t\,, & \mathrm{if\ }\lambda _{J,i}=0\,, \\
\frac{1-\exp (-2\Gamma \lambda _{J,i}(q)t)}{\lambda _{J,i}(q)}\,, & \mathrm{%
otherwise.}%
\end{array}%
\right.  \\
i& =1,\ldots ,4\,.
\end{split}
\label{eq:eigCrot}
\end{equation}%
We note that at least one eigenvalue of $J(q)$ equals zero by definition.

Finally, introduce a $4\times 4$-dimensional matrix $\sigma (t,q)$ such that
\begin{equation}  \label{lbec}
\sigma (t;q)\sigma^{\mathsf{T}} (t;q) = C(t;q).
\end{equation}
Since $C(t;q)$ is a symmetric matrix, $\sigma(t;q)$ can be determined as a
lower triangular matrix in the Cholesky decomposition of $C(t;q)$. LAPACK's
\texttt{dpotrf} can be used for this purpose.

With the above definitions, we obtain the following quasi-symplectic scheme
for (\ref{lt1})-(\ref{lt2}):

\begin{center}\underline{Langevin B}\end{center}
\begin{equation}
\begin{split}
\mathbf{P}_{0}& =\mathbf{p},\ \ \mathbf{R}_{0}=\mathbf{r},\ \mathbf{Q}_{0}=%
\mathbf{q},\ \ \mathbf{\Pi }_{0}=\mathbf{\pi ,} \\
\mathcal{P}_{1,k}& =\mathbf{P}_{k}e^{-\gamma h/2}+\sqrt{\frac{m}{\beta }%
(1-e^{-\gamma h})}\mathbf{\xi }_{k} \\
\mathit{\Pi }_{1,k}^{j}& =\exp \big(-\Gamma J(Q_{k}^{j})h/2\big) \Pi
_{k}^{j} \\
& +\sigma (h/2;Q_{k}^{j})\eta _{k}^{j},\ \ j=1,\ldots ,n,
\end{split}
\label{secla}
\end{equation}%
\begin{equation*}
\begin{split}
\mathcal{P}_{2,k}& =\mathcal{P}_{1,k}-\frac{h}{2}\nabla _{\mathbf{r}}U(%
\mathbf{R}_{k},\mathbf{Q}_{k}), \\
\mathit{\Pi }_{2,k}^{j}& =\mathit{\Pi }_{1,k}^{j}-\frac{h}{2}\nabla
_{q^{j}}U(\mathbf{R}_{k},\mathbf{Q}_{k}),\ \ j=1,\ldots ,n, \\
\mathbf{R}_{k+1}& =\mathbf{R}_{k}+\frac{h}{m}\mathcal{P}_{2,k},
\end{split}%
\end{equation*}%
\begin{equation*}
\begin{split}
(\mathcal{Q}_{1,k}^{j},\mathit{\Pi }_{3,k}^{j})& =\Psi _{3}(h/2;Q_{k}^{j},%
\mathit{\Pi }_{2,k}^{j}), \\
(\mathcal{Q}_{2,k}^{j},\mathit{\Pi }_{4,k}^{j})& =\Psi _{2}(h/2;\mathcal{Q}%
_{1,k}^{j},\mathit{\Pi }_{3,k}^{j}) \\
(\mathcal{Q}_{3,k}^{j},\mathit{\Pi }_{5,k}^{j})& =\Psi _{1}(h;\mathcal{Q}%
_{2,k}^{j},\mathit{\Pi }_{4,k}^{j}), \\
(\mathcal{Q}_{4,k}^{j},\mathit{\Pi }_{6,k}^{j})& =\Psi _{2}(h/2;\mathcal{Q}%
_{3,k}^{j},\mathit{\Pi }_{5,k}^{j}) \\
(Q_{k+1}^{j},\mathit{\Pi }_{7,k}^{j})& =\Psi _{3}(h/2;\mathcal{Q}_{4,k}^{j},%
\mathit{\Pi }_{6,k}^{j}),\ \ j=1,\ldots ,n,
\end{split}%
\end{equation*}%
\begin{eqnarray*}
\mathit{\Pi }_{8,k}^{j} &=&\mathit{\Pi }_{7,k}^{j}-\frac{h}{2}\nabla
_{q^{j}}U(\mathbf{R}_{k+1},\mathbf{Q}_{k+1}),\ j=1,\ldots ,n, \\
\mathcal{P}_{3,k} &=&\mathcal{P}_{2,k}-\frac{h}{2}\nabla _{\mathbf{r}}U(%
\mathbf{R}_{k+1},\mathbf{Q}_{k+1}),
\end{eqnarray*}%
\begin{equation*}
\begin{split}
\mathbf{P}_{k+1}& =\mathcal{P}_{3,k}e^{-\gamma h/2}+\sqrt{\frac{m}{\beta }%
(1-e^{-\gamma h})}\mathbf{\zeta }_{k},\ \  \\
\Pi _{k+1}^{j}& =\exp \big(-\Gamma J(Q_{k+1}^{j})h/2\big)\mathit{%
\Pi }_{8,k}^{j} \\
& +\sigma (h/2;Q_{k+1}^{j})\varsigma _{k}^{j},\ \ j=1,\ldots ,n, \\
k& =0,\ldots ,N-1,
\end{split}%
\end{equation*}%
where $\mathbf{\xi }_{k}=(\xi _{1,k},\ldots ,\xi _{3n,k})^{\mathsf{T}},$ $%
\mathbf{\zeta }_{k}=(\zeta _{1,k},\ldots ,\zeta _{3n,k})^{\mathsf{T}}$ and $%
\eta _{k}^{j}=(\eta _{1,k}^{j},\ldots ,\eta _{4,k}^{j})^{\mathsf{T}},$ $%
\varsigma _{k}^{j}=(\varsigma _{1,k}^{j},\ldots ,\varsigma _{4,k}^{j})^{%
\mathsf{T}},$ $j=1,\ldots ,n,$ with their components being i.i.d. with the
same law (\ref{n31}).

As in the case of the scheme (\ref{firla})-(\ref{n31}), the Jacobian $%
\mathbb{\bar{J}}$ of the one-step approximation corresponding to the
integrator (\ref{secla}), (\ref{n31}) is exactly equal to the Jacobian $%
\mathbb{J}$ of the original system (\ref{lt1})-(\ref{lt2}). The following
proposition can be proved.

\begin{proposition}
\label{prp2}The numerical scheme $(\ref{secla})$, $(\ref{n31})$ for $(\ref%
{lt1})$-$(\ref{lt2})$ is quasi-symplectic, it preserves the structural
property $(\ref{a211}),$ and is of weak order two.
\end{proposition}

We note that if we omit the rotational component in (\ref{lt1})-(\ref{lt2}),
the scheme (\ref{firla})-(\ref{n31}) coincides with a second-order weak
quasi-symplectic method from Ref.~\onlinecite{MT5} (see also Refs.~%
\onlinecite{MT6,MT7}) and the scheme (\ref{secla}), (\ref{n31}) is close to
the one from Ref.~\onlinecite{SKE99} (see also Refs.~%
\onlinecite{ISK01,MT5,MT6,Bus06}). Both stochastic integrators (\ref{firla}%
)-(\ref{n31}) and (\ref{secla}), (\ref{n31}) degenerate to the deterministic
scheme from Ref.~\onlinecite{qua02} when $\gamma =0$ and $\Gamma =0.$ The
scheme (\ref{secla}), (\ref{n31}) is usually preferable when $\gamma $
and/or $\Gamma $ are large (see our experimental results in Section~\ref%
{sec_tes} and a discussion in the case of translational Langevin equations
in Ref.~\onlinecite{Bus06}). It is slightly more expensive than (\ref{firla}%
)-(\ref{n31}) due to the need of generating the additional $7n$ random
variables $\zeta _{i,k}$ and $\varsigma _{k}^{j}$ per step and computing
Cholesky factorization. However, for most molecular system of practical
interest in computational chemistry and physics, where majority of the
computational effort is spent on force calculations, the additional cost is
negligible.

\subsection{Numerical scheme for the gradient-Langevin system\label%
{sec_numgrad}}

To construct the numerical scheme for the gradient-Langevin system (\ref{a10}%
)-(\ref{a100}), we exploit the Runge-Kutta method of order two for equations
with additive noise from Ref.~\onlinecite{MT6}[p. 113] to simulate the
\textquotedblleft gradient\textquotedblright\ part (\ref{a10}) and the
\textquotedblleft Langevin\textquotedblright\ rotational part (\ref{a100})
is approximated in the same way as in (\ref{secla}). The resulting
second-order weak scheme has the form%

\begin{center}\underline{gradient-Langevin}\end{center}
\begin{align}
\mathbf{R}_{0} &= \mathbf{r},\ \mathbf{Q}_{0}=\mathbf{q},\ \ \mathbf{\Pi }%
_{0}=\mathbf{\pi ,}  \label{firga} \\
\mathit{\Pi }_{1,k}^{j} &= \exp \big( -\Gamma J(Q_{k}^{j})h/2\big)
\Pi _{k}^{j}\notag +\sigma (h/2; Q_{k}^{j}) \eta_{k}^{j},\notag \\
j &= 1,\ldots ,n,  \notag
\end{align}%
\begin{eqnarray*}
\Delta R_{k} &=&-\frac{h}{2}\frac{\nu }{m}\nabla _{\mathbf{r}}U(\mathbf{R}%
_{k},\mathbf{Q}_{k})+\frac{\sqrt{h}}{2}\sqrt{\frac{2\nu }{m\beta }}\mathbf{%
\xi }_{k}, \\
\mathcal{R}_{k} &=&\mathbf{R}_{k}+2\times \Delta R_{k}, \\
\mathit{\Pi }_{2,k}^{j} &=&\mathit{\Pi }_{1,k}^{j}-\frac{h}{2}\nabla
_{q^{j}}U(\mathbf{R}_{k},\mathbf{Q}_{k}),\ j=1,\ldots ,n,
\end{eqnarray*}%
\begin{eqnarray*}
(\mathcal{Q}_{1,k}^{j},\mathit{\Pi }_{3,k}^{j}) &=&\Psi _{3}(h/2;Q_{k}^{j},%
\mathit{\Pi }_{2,k}^{j}), \\
(\mathcal{Q}_{2,k}^{j},\mathit{\Pi }_{4,k}^{j}) &=&\Psi _{2}(h/2;\mathcal{Q}%
_{1,k}^{j},\mathit{\Pi }_{3,k}^{j}) \\
(\mathcal{Q}_{3,k}^{j},\mathit{\Pi }_{5,k}^{j}) &=&\Psi _{1}(h;\mathcal{Q}%
_{2,k}^{j},\mathit{\Pi }_{4,k}^{j}), \\
(\mathcal{Q}_{4,k}^{j},\mathit{\Pi }_{6,k}^{j}) &=&\Psi _{2}(h/2;\mathcal{Q}%
_{3,k}^{j},\mathit{\Pi }_{5,k}^{j}) \\
(Q_{k+1}^{j},\mathit{\Pi }_{7,k}^{j}) &=&\Psi _{3}(h/2;\mathcal{Q}_{4,k}^{j},%
\mathit{\Pi }_{6,k}^{j}),\ \ j=1,\ldots ,n,
\end{eqnarray*}%
\begin{eqnarray*}
\mathbf{R}_{k+1} &=&\mathcal{R}_{k}-\Delta R_{k}+\frac{\sqrt{h}}{2}\sqrt{%
\frac{2\nu }{m\beta }}\mathbf{\xi }_{k} \\
&&-\frac{h}{2}\frac{\nu }{m}\nabla _{\mathbf{r}}U(\mathcal{R}_{k},\mathbf{Q}%
_{k+1}), \\
\mathit{\Pi }_{8,k}^{j} &=&\mathit{\Pi }_{7,k}^{j}-\frac{h}{2}\nabla
_{q^{j}}U(\mathbf{R}_{k+1},\mathbf{Q}_{k+1}),
\end{eqnarray*}%
\begin{align*}
\Pi _{k+1}^{j} &= \exp \big(-\Gamma J(Q_{k+1}^{j})h/2\big)
\mathit{\Pi }_{8,k}^{j}+\sigma (h/2; Q_{k+1}^{j})
\varsigma _{k}^{j}, \\
j &= 1,\ldots ,n;\ \ k=0,\ldots ,N-1,
\end{align*}%
where $\mathbf{\xi }_{k}=(\xi _{1,k},\ldots ,\xi _{3n,k})^{\mathsf{T}}$ and $%
\eta _{k}^{j}=(\eta _{1,k}^{j},\ldots ,\eta _{4,k}^{j})^{\mathsf{T}},$ $%
\varsigma _{k}^{j}=(\varsigma _{1,k}^{j},\ldots ,\varsigma _{4,k}^{j})^{%
\mathsf{T}},$ $j=1,\ldots ,n,$ with their components being i.i.d. with the
same law (\ref{n31}). The following proposition can be proved.

\begin{proposition}
\label{prp3}The numerical scheme $(\ref{firga})$, $(\ref{n31})$ for $(\ref%
{a10})$-$(\ref{a100})$ preserves the structural property $(\ref{a211})$ and
is of weak order two.
\end{proposition}

We draw attention to the fact that the above gradient-Langevin scheme requires two force
calculations per step and thus is approximately twice as expensive as the
Langevin schemes presented in Section \ref{sec_numla}.

\subsection{Computational errors\label{sec_err}}

Let us recall that the objective is to compute highly multi-dimensional
integrals with respect to the Gibbsian measure $\mu (x)$ with the density (%
\ref{a3}). The considered stochastic systems (\ref{lt1})-(\ref{lt2}) and (%
\ref{a10})-(\ref{a100}) are assumed to be ergodic with the Gibbsian
invariant measure and we can represent the integrals of interest as (cf. (%
\ref{PA31})):%
\begin{equation}
\varphi ^{erg}=\int \varphi (x)\,d\mu (x)=\lim_{t\rightarrow \infty
}E\varphi (X(t;x)).  \label{PAX}
\end{equation}%
We are interested here in systems solutions of which satisfy a stronger
condition, namely they are exponentially ergodic, i.e., for any $x\in
\mathbb{R}^{14n}$ and any function $\varphi$ with a polynomial growth:
\begin{equation}
\left\vert E\varphi (X(t;x))-\varphi ^{erg}\right\vert \leq Ce^{-\lambda
t},\ \ t\geq 0,  \label{PA34}
\end{equation}%
where $C>0$ and $\lambda >0$ are some constants. In Refs.~%
\onlinecite{Soize,Stua,Tal02} (see also references therein), one can find
conditions under which Langevin equations are exponentially ergodic.

It follows from (\ref{PA34}) (and (\ref{PAX})) that for any $\varepsilon >0$
there exists $T_{0}>0$ such that for all $T\geq T_{0}$
\begin{equation}
\left\vert E\varphi (X(T;x))-\varphi ^{erg}\right\vert \leq \varepsilon .
\label{PA35}
\end{equation}%
Then we can use the following estimate for the ergodic limit $\varphi ^{erg}$%
:
\begin{eqnarray}
\varphi ^{erg} &\approx &E\varphi (X(T;x))\approx E\varphi (\bar{X}(T;x))
\label{S1} \\
&\approx &\hat{\varphi}^{erg}:=\frac{1}{L}\sum_{l=1}^{L}\varphi \left( \bar{X%
}^{(l)}(T;x)\right) ,  \notag
\end{eqnarray}%
where $T$ is a sufficiently large time, $\bar{X}$ is an approximation of $X,$
and $L$ is the number of independent approximate realizations. The total
error
\begin{equation}
R_{\hat{\varphi}^{erg}}:=\hat{\varphi}^{erg}-\varphi ^{erg}  \label{S2}
\end{equation}%
consists of three parts: the error $\varepsilon $ of the approximation $%
\varphi ^{erg}$ by $E\varphi (X(T;x))$; the error of numerical integration $%
Ch^{p}$ (see (\ref{A20})), and the Monte Carlo error; i.e.,
\begin{equation*}
R_{\hat{\varphi}^{erg}}\sim Ch^{p}+\varepsilon +O\left( \frac{1}{\sqrt{L}}%
\right) ,
\end{equation*}%
or more specifically
\begin{equation*}
Bias(\hat{\varphi}^{erg})=\left\vert E\hat{\varphi}^{erg}-\varphi
^{erg}\right\vert \leq Ch^{p}+\varepsilon ,
\end{equation*}%
\begin{equation*}
Var(\hat{\varphi}^{erg})=O(1/L).
\end{equation*}%
Each error is controlled by its own parameter: sufficiently large $T$
ensures smallness of the error $|\varphi ^{erg}-E\varphi (X(T;x))|;$ time
step $h$ (as well as the choice of numerical method) controls the
numerical integration error; the statistical error is regulated by choosing
an appropriate number of independent trajectories $L.$

The other, commonly used in molecular dynamics, numerical\emph{\ }approach
to calculating ergodic limits is based on the known equality
\begin{equation}
\lim_{t\rightarrow \infty }\frac{1}{t}\int\limits_{0}^{t}\varphi
(X(s;x))ds=\varphi ^{erg}\ \ a.s.,  \label{PB51}
\end{equation}%
where the limit does not depend on $x.$ Then by approximating a single
trajectory, one gets the following estimator for $\varphi ^{erg}$:%
\begin{equation}
\varphi ^{erg}\sim \frac{1}{\tilde{T}}\int\limits_{0}^{\tilde{T}}\varphi
(X(s;x))ds\sim \check{\varphi}^{erg}:=\frac{1}{L}\sum_{l=1}^{L}\varphi (\bar{%
X}(lh;x)),  \label{PB52}
\end{equation}%
where $\tilde{T}$ is sufficiently large and $Lh=\tilde{T}.$ In Ref.~%
\onlinecite{Tal90} this approach was rigorously justified in the case of
ergodic SDEs with nondegenerate noise and globally Lipschitz coefficients.
Let us emphasize that $\tilde{T}$ in (\ref{PB52}) is much larger than $T$ in
(\ref{PA35}) and (\ref{S1}) because $\tilde{T}$ should be such that it not
just ensures the distribution of $X(t)$ to be close to the invariant
distribution (like it is required from $T)$ but it should also guarantee
smallness of variance of $\check{\varphi}^{erg}$. See further details
concerning computing ergodic limits in Ref.~\onlinecite{MT7} and references
therein.

\section{Numerical Investigation\label{sec_tes}}

In this section we present a numerical study of the Langevin and
gradient-Langevin thermostats derived in Section~\ref{sec_mot}. In
particular, we investigate the dependence of the thermostat properties on
the choice of parameters $\gamma $ and $\Gamma $ for the Langevin system (%
\ref{lt1})-(\ref{lt2}) and $\nu $ and $\Gamma $ for the gradient-Langevin
system (\ref{a10})-(\ref{a100}), as well as the dependence of the numerical
discretization errors of the numerical schemes Langevin A, Langevin B, and
gradient-Langevin on the integration step size $h$. As a model system, we
use the popular TIP4P rigid model of water\cite{Jorgensen83}. In order to
speed up the simulations, both Lennard-Jones and electrostatic interactions
are smoothly turned off between $9.5$ and $10$\thinspace \AA . This
truncation has minimal effect on the structure of liquid water, but leads to
a lower estimated melting temperature\cite{Handel08} of $219$\thinspace K.


The two key requirements of a thermostat are: i) correct sampling of phase
space points distributed according to the Gibbs distribution at a desired
thermostat temperature $T$, and ii) rapid relaxation of the system to the
desired equilibrium state. The numerical accuracy of the sampling can be
estimated by comparing the values of various system properties (e.g.
kinetic and potential energies, pressure) averaged over long simulation
runs to those obtained with a much smaller step size $h$.

To estimate how quickly the system relaxes to the desired equilibrium state
we use the following simple experiment. A system of 2000 TIP4P water
molecules is equilibrated at $T_{0}=220$\thinspace K. Then the temperature
of the thermostat is increased (instantaneously) to $T_{1}=270$\thinspace K,
and the run is continued until the system is equilibrated at the new
temperature. We deliberately choose to simulate the system at lower
temperatures (close to the melting temperature for this model of water),
where the relaxation of the system is expected to be slower.

Assuming that the system is exponentially ergodic (see (\ref{PA34})), we can
expect that any measured quantity $A$ will relax from its equilibrium value $%
A_{0}$ at $T_{0}$ to the equilibrium value $A_{1}$ at $T_{1}$ according to
the approximate formula
\begin{equation}
EA(t)\equiv \langle A(t)\rangle \doteq A_{1}+(A_{0}-A_{1})\exp (-t/\tau
_{A})\,,  \label{eq:expfun}
\end{equation}%
where $\tau _{A}$ is the characteristic relaxation time of the quantity $A$.
The temperature switch occurs at $t=0$ and the angle brackets denote average
over an ensemble of independent simulation runs. The subscript on $\tau _{A}$
indicates that different quantities may relax with different rates. The rate
of system equilibration should be estimated from the \emph{maximum} value of
$\tau _{A}$ among the quantities of interest.

The quantities we measure include the translational kinetic temperature
\begin{equation}  \label{eq:Ttr}
\mathcal{T}_\mathrm{tr} = \frac{\mathbf{p}^{\mathsf{T}} \mathbf{p}} {3nk_B m}%
\,,
\end{equation}
rotational kinetic temperature
\begin{equation}  \label{eq:Trot}
\mathcal{T}_\mathrm{rot} = \frac{2}{3nk_B }\sum_{j=1}^n%
\sum_{l=1}^3V_{l}(q^j,\pi^j)\,,
\end{equation}
and potential energy per molecule
\begin{equation}  \label{eq:Pot}
\mathcal{U} = \frac{1}{n}U(\mathbf{r},\mathbf{q})\,.
\end{equation}

\begin{figure}[tbp]
\includegraphics[width=\figurewidth]{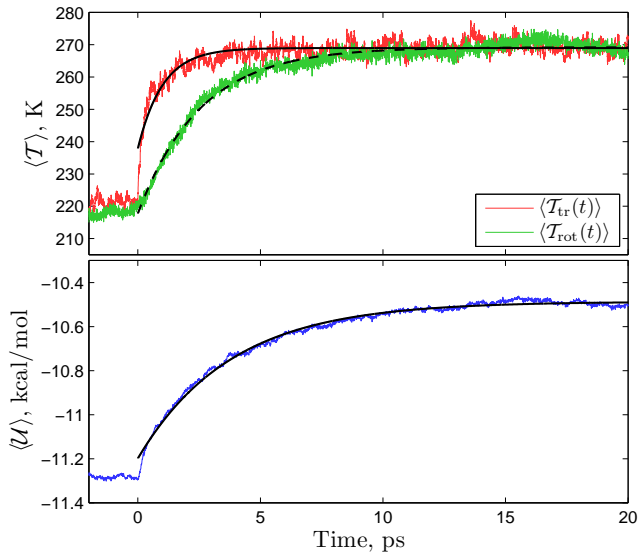}
\caption{(Color online) Relaxation dynamics with translational Langevin thermostat:
$\gamma = 4.0\,$ps$^{-1}$, $\Gamma = 0$. Thin lines show relaxation
dynamics averaged over ten independent runs, thick lines (solid and dashed
for translational and rotational temperatures, respectively) show the least
squares fit to formula (\protect\ref{eq:expfun}).  The estimated values of the
relaxation times are $\tau_{\mathcal{T}_\mathrm{tr}} = 0.2\,$ps,
$\tau_{\mathcal{T}_\mathrm{rot}} = 1.9\,$ps, and
$\tau_\mathcal{U} = 3.6\,$ps.}
\label{fig:relax-lin}
\end{figure}

To illustrate the response of the system to the instantaneous temperature
change, we show in Fig.~\ref{fig:relax-lin} the result of applying the
Langevin thermostat only to the translational degrees of freedom, i.e. $%
\Gamma = 0$ in (\ref{lt1})-(\ref{lt2}). As expected, the translational
kinetic temperature quickly relaxes to the new temperature, while the
rotational kinetic temperature and potential energy lag behind. To estimate
the relaxation rates of the measured quantities, we use the least squares
fit of the exponential function (\ref{eq:expfun}) to the average measured
quantity $\langle A(t) \rangle$.

In all the simulations we performed, the potential energy relaxation time is
larger than that for either of the kinetic temperatures. Therefore, we
determine the relaxation time of the system to the new equilibrium state
based on the value of $\tau_{\mathcal{U}}$.

\begin{figure}[tbp]
\includegraphics[width=\figurewidth]{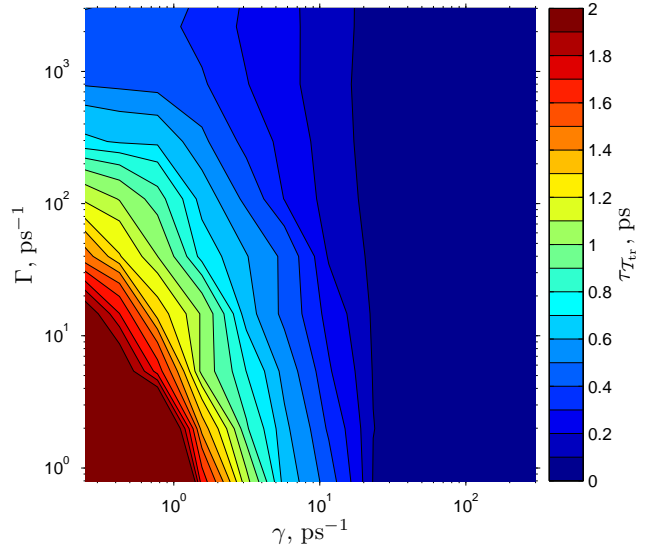}
\caption{(Color online) Translational temperature relaxation time for the Langevin
thermostat {(\protect\ref{lt1})-(\protect\ref{lt2})} as a function of the
thermostat parameters $\protect\gamma$ and $\Gamma$.}
\label{fig:tauT}
\end{figure}

\begin{figure}[tbp]
\includegraphics[width=\figurewidth]{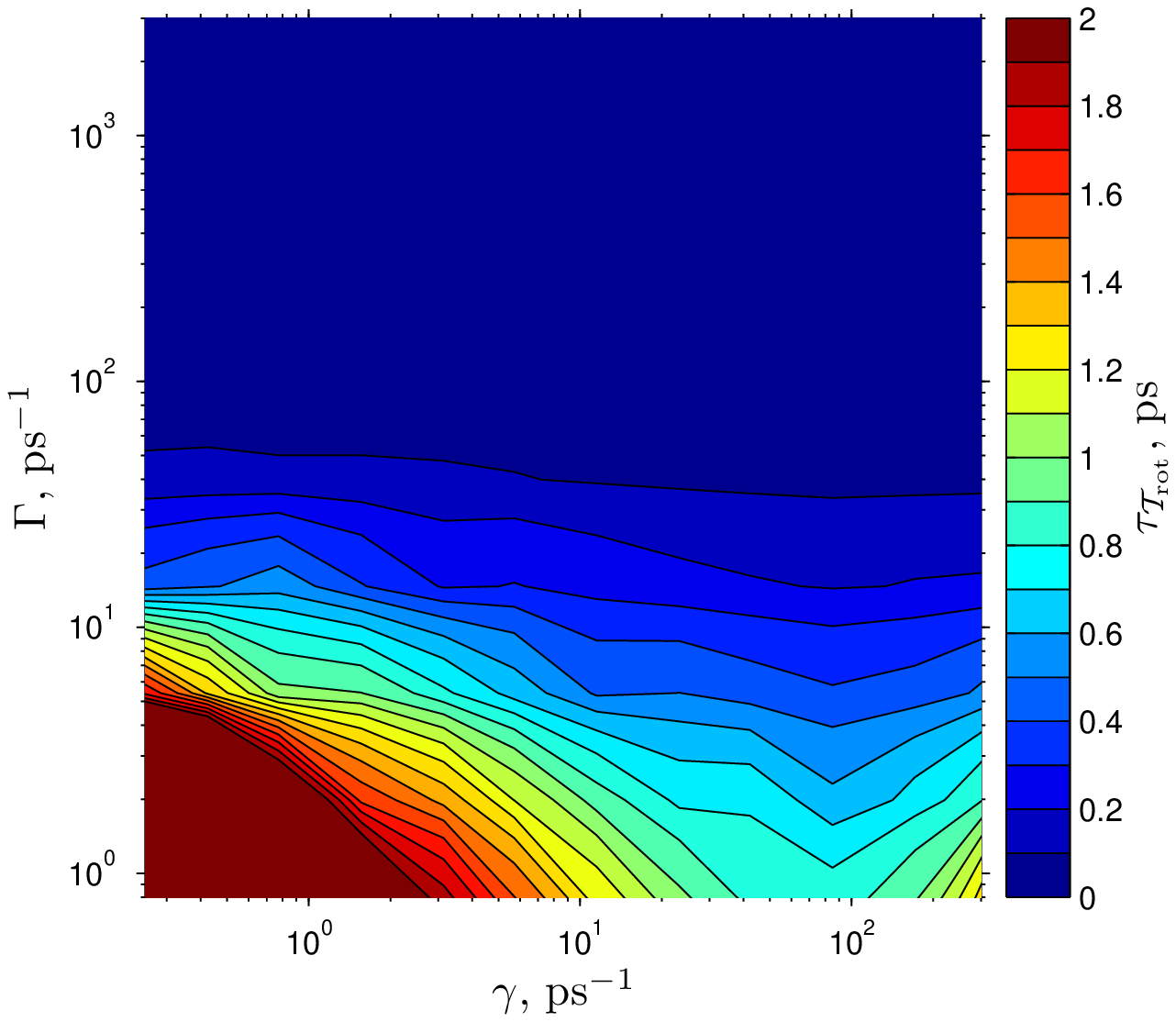}
\caption{(Color online) Rotational temperature relaxation time for the Langevin
thermostat {(\protect\ref{lt1})-(\protect\ref{lt2})} as a function of
the thermostat parameters $\protect\gamma$ and $\Gamma$.}
\label{fig:tauR}
\end{figure}

\begin{figure}[tbp]
\includegraphics[width=\figurewidth]{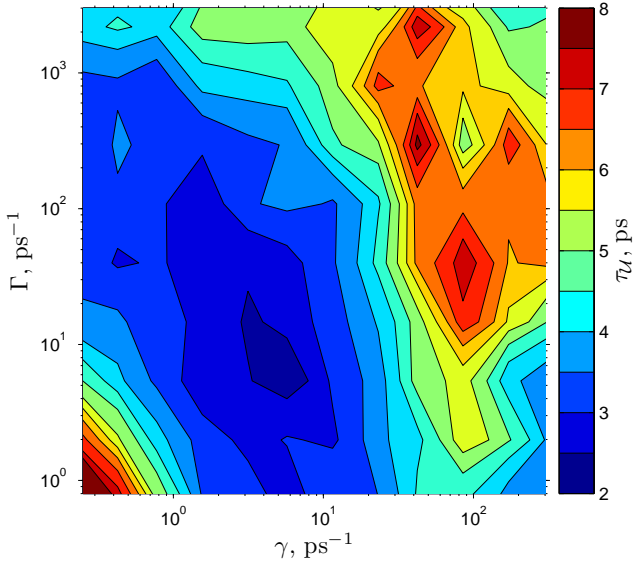}
\caption{(Color online) Potential energy relaxation time for the Langevin
thermostat {(\protect\ref{lt1})-(\protect\ref{lt2})} as a function of the
thermostat parameters $\protect\gamma$ and $\Gamma$.}
\label{fig:tauU}
\end{figure}

Varying the value of the translational Langevin parameter $\gamma $, we
observe that relaxation is slower for both small and large values of $\gamma
$, with the fastest relaxation around $\gamma = 4.0\,$ps$^{-1}$. The
existence of an optimal value for the choice of the thermostat parameter is
consistent with observations in Ref.~\onlinecite{Bus06} and can be
understood in terms of the interaction of the system with the thermostat.
For small values of $\gamma $, the relaxation of the system is slow due to
the limited heat flux between the system and the thermostat. For large $%
\gamma $, even though the kinetic temperature relaxes very quickly, the
relaxation of the configurational state of the system is apparently hindered
by the disruptive influence of the random force on the Hamiltonian dynamics
which is driving the system to the new equilibrium.

\begin{figure}[tbp]
\includegraphics[width=\figurewidth]{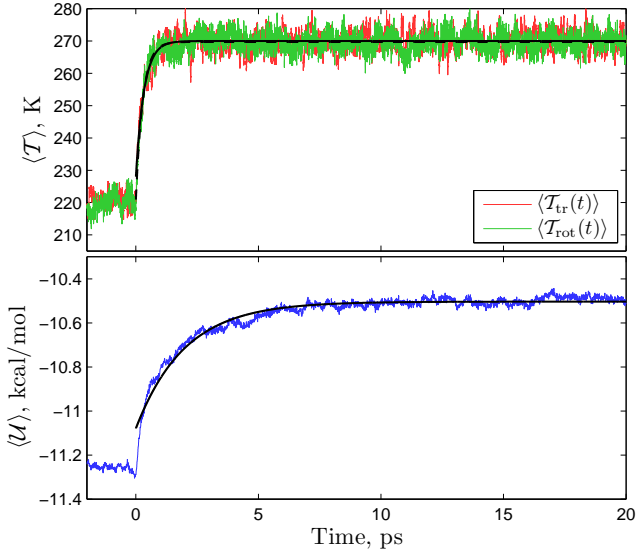}
\caption{(Color online) Relaxation dynamics with `optimal' choice of Langevin
thermostat parameters: $\protect\gamma =4.0\,$ps$^{-1}$, $\Gamma = 10.0\,$ps$^{-1}$.
Thin lines show relaxation dynamics averaged over ten independent runs,
thick lines (solid and dashed for translational and rotational temperatures,
respectively) show the least squares fit to formula (\protect\ref{eq:expfun}).
The estimated values of the relaxation times are
$\tau_{\mathcal{T}_\mathrm{tr}} = 0.28\,$ps,
$\tau_{\mathcal{T}_\mathrm{rot}} = 0.26\,$ps, and
$\tau_\mathcal{U} = 2.0\,$ps.}
\label{fig:relax-opt}
\end{figure}

\subsection{Langevin Thermostats}

Next, we investigate the dependence of the system relaxation time $\tau _{%
\mathcal{U}}$ on both $\gamma $ and $\Gamma $ in (\ref{lt1})-(\ref{lt2}). In
order to minimize the influence of the numerical discretization error, we
use a relatively small time step of $0.2$\thinspace fs. With such a small
time step, the difference between Langevin A and Langevin B is negligible
compared to the sampling error. To produce the results reported below, we
use Langevin A. We evaluate $\tau _{\mathcal{U}}$ on a logarithmic grid of $%
\gamma $ and $\Gamma $ values using five independent runs at each point. The
results are shown in Figs.~\ref{fig:tauT}, \ref{fig:tauR}, and \ref{fig:tauU}%
. As expected, the relaxation speed of translational and rotational
temperatures uniformly increases with increasing values of $\gamma $ and $%
\Gamma $, respectively. At the same time, the relaxation speed of the
potential energy exhibits nonuniform dependence on the thermostat
parameters. As can be seen in Fig.~\ref{fig:tauU}, the fastest relaxation of
the system is achieved when the Langevin thermostat is applied to both
translational and rotational degrees of freedom, with $\gamma =2-8$%
\thinspace ps$^{-1}$ and $\Gamma =3-40$\thinspace ps$^{-1}$. The relaxation
dynamics of the system with `optimal' choice of Langevin thermostat
parameters is demonstrated in Fig.~\ref{fig:relax-opt}. In this case $\tau _{%
\mathcal{T}_{\mathrm{tr}}}=0.28\,$ps, $\tau _{\mathcal{T}_{\mathrm{rot}%
}}=0.26\,$ps, and $\tau _{\mathcal{U}}=2.0\,$ps, which shows that the system
relaxation is almost twice as fast as when the Langevin thermostat is
applied only to translational degrees of freedom. Note that the results
shown in Fig.~\ref{fig:tauU} for $\gamma $ larger than about $100\,$ps$^{-1}$ or
$\Gamma$ larger than about $1000\,$ps$^{-1}$ are not
reliable due to excessive coupling of the system to the thermostat, which
disrupts the Hamiltonian flow of the system. In this case, the relaxation
dynamics is poorly represented by the exponential function (\ref{eq:expfun})
and thus the fits produce misleading values for the system relaxation time.

\begin{figure}[tbp]
\includegraphics[width=\figurewidth]{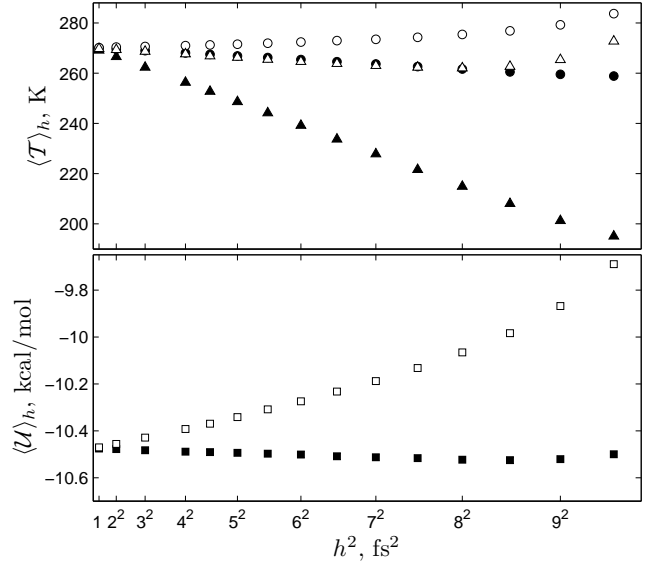}
\caption{Dependence of the approximated average properties of a system of
2000 TIP4P water molecules on the integration time step $h$ for Langevin A
and B. The system is equilibrated with the thermostat parameters $\protect%
\gamma =4.0$\thinspace ps$^{-1}$, $\Gamma =10.0$\thinspace ps$^{-1}$, and $%
T=270$\thinspace K. The quantities $\langle \mathcal{T}_{\mathrm{tr}}\rangle
_{h}$, $\langle \mathcal{T}_{\mathrm{rot}}\rangle _{h}$, and $\langle
\mathcal{U}\rangle _{h}$ are denoted by circles, triangles, and squares,
respectively. Solid and open symbols refer to Langevin A and B, respectively.}
\label{fig:error}
\end{figure}

Now we look at performance of the numerical integrators proposed in Section~%
\ref{sec_numla}\ for the Langevin thermostat (\ref{lt1})-(\ref{lt2}). Since
both Langevin A and B are second-order methods, the calculated average
quantities for simulations with step size $h$ should have the form\cite%
{TAT90,MT6}
\begin{equation}
  \langle A\rangle _{h}=\langle A\rangle _{0}+C_{A}h^{2}+O(h^{3})\,,
\label{eq:conv}
\end{equation}%
where $\langle A\rangle _{h}$ denotes the average value of dynamical
quantity $A(t)$ calculated over a numerical trajectory with time step $h$.
The dependence of $\langle \mathcal{T}_{\mathrm{tr}}\rangle _{h}$, $\langle
\mathcal{T}_{\mathrm{rot}}\rangle _{h}$, and $\langle \mathcal{U}\rangle _{h}
$ on $h$ for both Langevin A and B is illustrated in Fig.~\ref{fig:error}.
It appears that Langevin A has larger discretization error for the
rotational temperature and smaller error for the potential energy than
Langevin B. The linear dependence of the measured quantities on $h^{2}$ is
maintained up to a relatively large time step of about $h=7\,$fs.
The values of the slopes $C_{A}$ are listed in Table~\ref{tab:error}. Both
methods become unstable at about $h=10\,$fs.

\begin{table}[tbp]
\caption{Values of the coefficients $C_{A}$ in the discretization errors (%
\protect\ref{eq:conv}) for the measured quantities with Langevin A and B.
The system's parameters are as in Fig.~\protect\ref{fig:error}.}%
\begin{ruledtabular}
\begin{tabular}{l|cc}
& Langevin A & Langevin B \\\hline
$C_{\mathcal{T}_\mathrm{tr}}$\,,~K/fs$^{2}$  & $-0.13$   & $0.06$ \\
$C_{\mathcal{T}_\mathrm{rot}}$\,,~K/fs$^{2}$ & $-0.85$   & $-0.14$ \\
$C_{\mathcal{U}}$\,,~kcal/mol/fs$^{2}$       & $-0.0007$ & $0.0056$
\end{tabular}
\end{ruledtabular}
\label{tab:error}
\end{table}

\begin{figure}[tbp]
\includegraphics[width=\figurewidth]{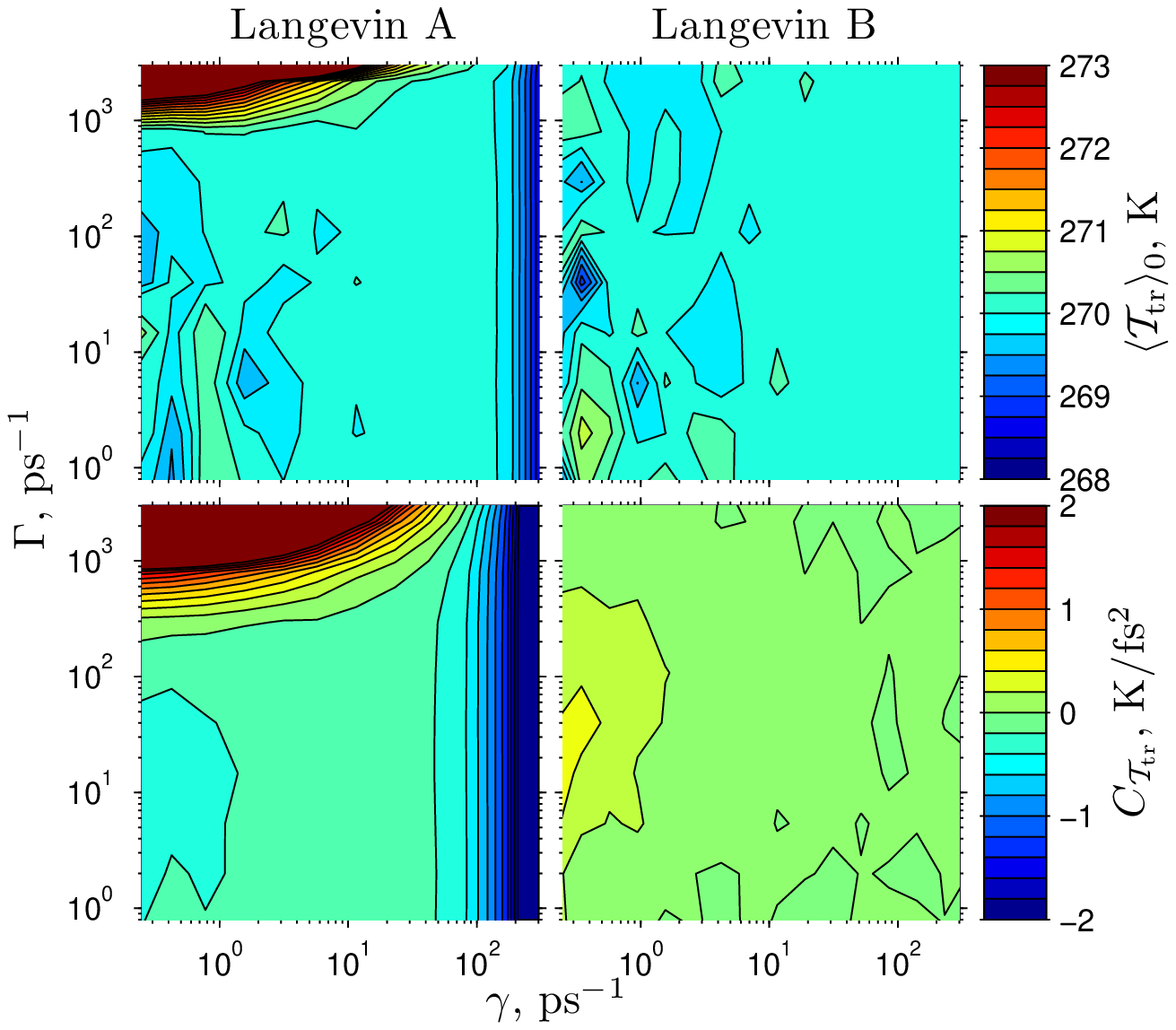}
\caption{(Color online) Dependence of $\langle\mathcal{T}_\mathrm{tr}\rangle_0$
and $C_{\mathcal{T}_\mathrm{tr}}$ on $\protect\gamma$ and $\Gamma$
for Langevin A and B thermostats.}
\label{fig:errorTtr}
\end{figure}

\begin{figure}[tbp]
\includegraphics[width=\figurewidth]{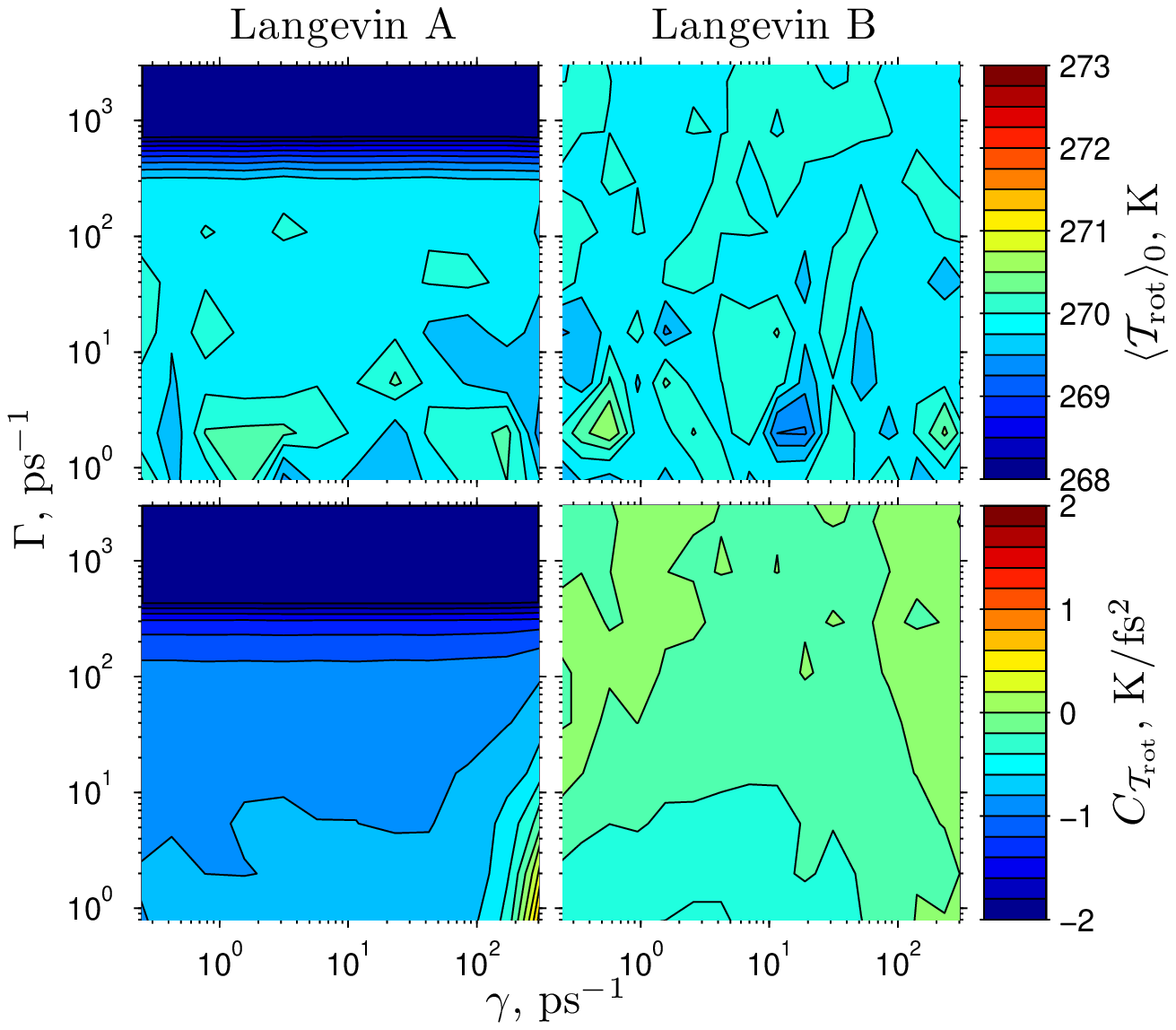}
\caption{(Color online) Dependence of $\langle\mathcal{T}_\mathrm{rot}\rangle_0$
and $C_{\mathcal{T}_\mathrm{rot}}$ on $\protect\gamma$ and $\Gamma$
for Langevin A and B thermostats.}
\label{fig:errorTrot}
\end{figure}

\begin{figure}[tbp]
\includegraphics[width=\figurewidth]{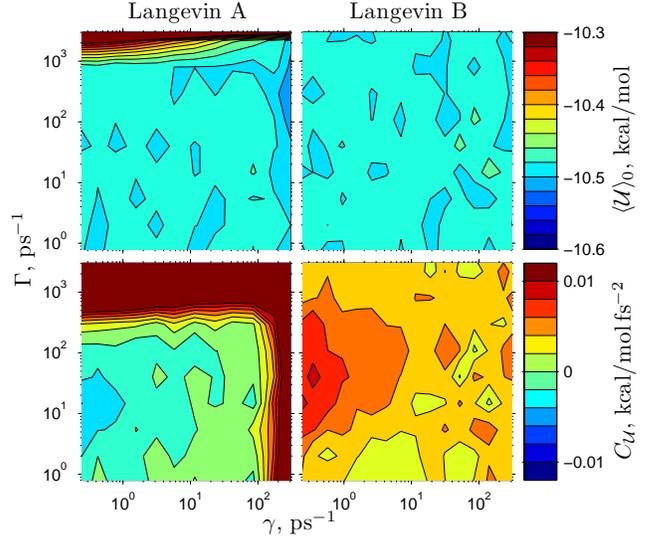}
\caption{(Color online) Dependence of $\langle\mathcal{U}\rangle_0$
and $C_{\mathcal{U}}$ on $\protect\gamma$ and $\Gamma$
for Langevin A and B thermostats.}
\label{fig:errorU}
\end{figure}

We have investigated the dependence of $\langle A\rangle _{0}$ and $C_{A}$
on the thermostat parameters $\gamma $ and $\Gamma $, by running simulations
with time steps $h=2\,$fs and $3\,$fs and estimating these
quantities from the straight line fit with respect to $h^{2}$. We consider
the fit to be justified if the higher order terms in (\ref{eq:conv}) are
small, i.e., when the quantities $\langle \mathcal{T}_{\mathrm{tr}}\rangle
_{0}$ and $\langle \mathcal{T}_{\mathrm{rot}}\rangle _{0}$ determined from (%
\ref{eq:conv}) are equal to the thermostat temperature parameter $T=270\,$K.
In Fig.~\ref{fig:errorTtr} we show results for the translational temperature
measurements in simulations with Langevin A and B. As expected, $\langle
\mathcal{T}_{\mathrm{tr}}\rangle _{0}$ converges to $270\,$K. We
note in passing that smaller statistical errors are observed at larger
values of $\gamma $. The behavior of $C_{\mathcal{T}_{\mathrm{tr}}}$ for
Langevin A exhibits a plateau for small and moderate values of $\gamma $ and
$\Gamma $ and then changes rapidly at values that are `too large' for this
system.  By contrast, the translational temperature discretization error of
Langevin B thermostat exhibits consistent behavior for all values of $\gamma
$ and $\Gamma $. Similar differences between Langevin A and B can be seen in
the measurements of rotational temperature and potential energy shown in
Figs.~\ref{fig:errorTrot} and \ref{fig:errorU}, respectively. As can be seen
from the plot of $\langle \mathcal{T}_{\mathrm{rot}}\rangle _{0}$ in Fig.~%
\ref{fig:errorTrot}, the straight line fit also breaks down at large $\Gamma
$ values in Langevin A.



\subsection{Gradient-Langevin thermostat}

Here we describe numerical experiments with the gradient-Langevin thermostat
(\ref{a10})-(\ref{a100}) introduced in Section~\ref{sec_grad}. Since the
gradient system for the translational motion does not include linear momenta
$\mathbf{P}$, the translational kinetic temperature $\mathcal{T}_{\mathrm{tr}%
}$ is not available for measurement in this case. Therefore, in our
numerical experiments we measure the rotational temperature $\mathcal{T}_{%
\mathrm{rot}}$ and potential energy per particle $\mathcal{U}$ as defined by
(\ref{eq:Trot}) and (\ref{eq:Pot}), respectively.

For the particular system studied here, the gradient-Langevin numerical
scheme (\ref{firga}), (\ref{n31}) from Section~\ref{sec_numgrad}\ becomes
unstable when the product $h\nu $ is larger than about $200\,$fs$^{2}$.
Therefore, with the step size of $h=0.2\,$fs used in our simulations,
we can study the properties of the gradient-Langevin scheme with $\nu $ up
to about $1000\,$fs.

\begin{figure}[tbp]
\includegraphics[width=\figurewidth]{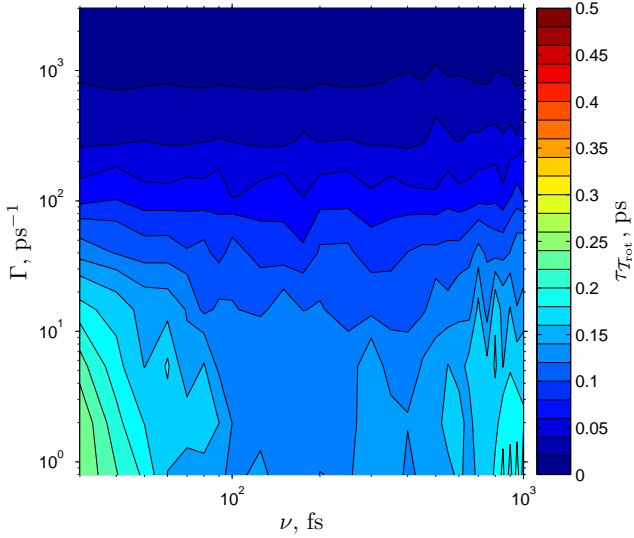}
\caption{(Color online) Rotational temperature relaxation time for the
gradient-Langevin thermostat {(\protect\ref{a10})-(\protect\ref{a100})}
as a function of the thermostat parameters $\protect\nu$ and $\Gamma$.}
\label{fig:tauR-grad}
\end{figure}

\begin{figure}[tbp]
\includegraphics[width=\figurewidth]{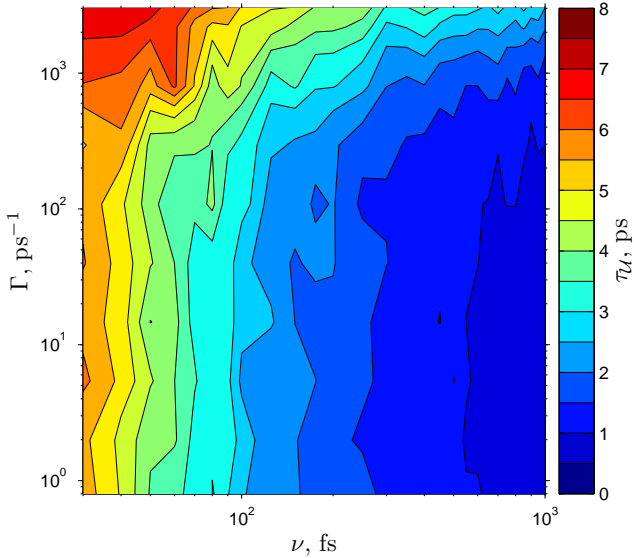}
\caption{(Color online) Potential energy relaxation time for the
gradient-Langevin thermostat {(\protect\ref{a10})-(\protect\ref{a100})}
as a function of the thermostat parameters $\protect\nu$ and $\Gamma$.}
\label{fig:tauU-grad}
\end{figure}

\begin{figure}[tbp]
\includegraphics[width=\figurewidth]{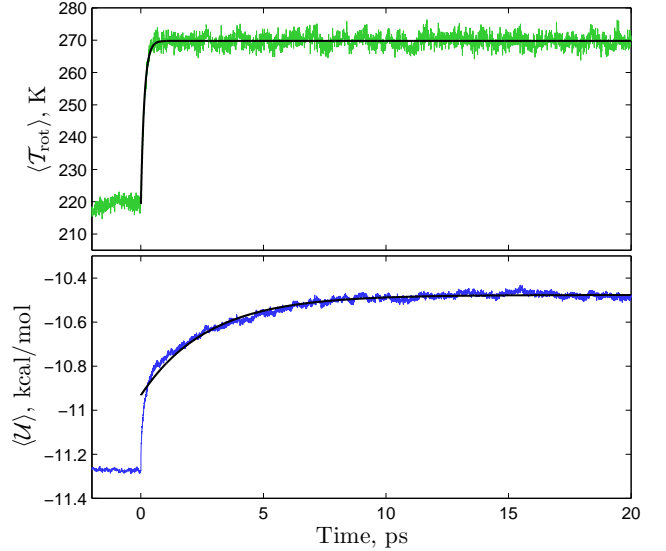}
\caption{(Color online) Relaxation dynamics with gradient-Langevin thermostat:
$\protect\nu = 100\,$fs, $\Gamma = 0$. Thin lines show relaxation dynamics
averaged over seven independent runs, thick lines show the least squares
fit to formula (\protect\ref{eq:expfun}).}
\label{fig:relax-grad}
\end{figure}

We conducted the relaxation experiment, where we monitored $\langle \mathcal{%
T}_{\mathrm{rot}}(t)\rangle $ and $\langle \mathcal{U}(t)\rangle $ while the
thermostat temperature parameter was switched from 220\thinspace K to
270\thinspace K. The relaxation times $\tau _{\mathcal{T}_{\mathrm{rot}}}$
and $\tau _{\mathcal{U}}$ for these quantities were calculated for different
values of $\nu $ and $\Gamma $. The results are shown in Figs.~\ref%
{fig:tauR-grad} and \ref{fig:tauU-grad}. As expected, the relaxation time
for rotational temperature decreases with increasing value of $\Gamma $. The
somewhat surprising finding of this experiment is that, even for small
values of $\Gamma $ the relaxation time is much smaller here than in the
case of Langevin thermostat (see Fig.~\ref{fig:tauR}). As an illustration,
we show the relaxation experiment for $\nu =100\,$fs and $\Gamma =0$ in Fig.~%
\ref{fig:relax-grad}. The estimated relaxation time for
rotational temperature, $\tau _{\mathcal{T}_{\mathrm{rot}}}=0.14\,$ps, is
much smaller then the corresponding quantity for the Langevin thermostat,
even though the relaxation time for the potential energy, $\tau _{\mathcal{U}%
}=2.7\,$ps, is similar. This indicates a very efficient heat transfer
between the gradient dynamics of the translational motion and the rotational
motion.

The dependence of $\tau _{\mathcal{U}}$ on the thermostat parameters for the
gradient-Langevin system shown in Fig.~\ref{fig:tauU-grad} is markedly
different than for the Langevin system. In particular, the relaxation time
decreases with increasing $\nu $ without reaching a minimum value within the
range of $\nu $ values explored. At the same time, there is little
dependence on $\Gamma $, except for very large $\Gamma $ where, similar to
the Langevin system, the measurements of the relaxation time are not
reliable. Also, note that the relaxation times are much smaller, reaching as
low as $0.7\,$ps for $\nu =1000\,$fs, compared to the minimum value of about
$2.0\,$ps for the Langevin system.

Of course, the direct comparison between the relaxation speeds of
gradient-Langevin and Langevin dynamics has to be taken with caution, since,
as we mentioned in Section \ref{sec_grad}, the gradient dynamics does not
have a natural evolution time. In particular, the mass of the molecule $m$
does not have a specific meaning in the gradient system, since it can be
rescaled to any value together with $h$ and $\nu $ (see (\ref{a10})).
Computationally, the relaxation speed depends on the time step $h$ and,
while with $\nu =1000\,$fs the gradient Langevin scheme becomes unstable for
$h$ larger than $0.2$\thinspace fs, the Langevin scheme remains stable up to
about $h=10\,$fs for the optimal values of $\gamma =4.0\,$ps$^{-1}$ and $%
\Gamma =10\,$ps$^{-1}$.

\begin{figure}[tbp]
\includegraphics[width=\figurewidth]{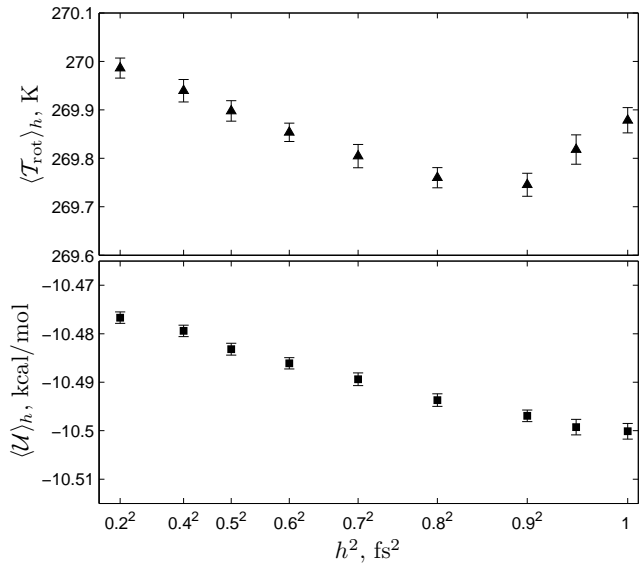}
\caption{Dependence of the approximated average properties of
a system of 2000 TIP4P water molecules on the integration time step $h$ for the
gradient-Langevin numerical method. The system is equilibrated with the
thermostat parameters $\protect\nu =200\,$fs, $\Gamma =5.0\,$ps$^{-1}$, and
$T=270\,$K.  The quantities $\langle\mathcal{T}_{\mathrm{rot}}\rangle _{h}$,
and $\langle \mathcal{U}\rangle _{h}$ are denoted by circles and squares,
respectively. Error bars reflect 95\% confidence intervals in the obtained
results estimated from block averages.}
\label{fig:error_grad}
\end{figure}

The dependence of discretization error in measured quantities on $h$ for the
gradient-Langevin scheme is shown in Fig.~\ref{fig:error_grad}.  As in the
case of Langevin A and B, we clearly see the linear dependence of
$\langle\mathcal{T}_{\mathrm{rot}}\rangle _{h}$, and
$\langle \mathcal{U}\rangle _{h}$ on $h^2$.  The estimated slopes
in (\ref{eq:conv}) are
$C_{\mathcal{T}_\mathrm{rot}} = -0.38\,$K/fs$^{2}$ and
$C_{\mathcal{U}} = -0.029\,$kcal/mol/fs$^{2}$.
Unfortunately, for this value of $\nu$ the gradient-Langevin numerical
integrator becomes unstable for $h>1\,$fs, which is rather small,
given that the Langevin A and B integrator are stable for $h$ up to about
10\,fs.  Still, given the observed efficient heat transfer from the gradient
subsystem for translational dynamics to the rotational dynamics (see
Fig.~\ref{fig:relax-grad} and related discussion), it might be of
interest to construct numerical methods for the gradient-Langevin
system (\ref{a10})-(\ref{a100}) with better stability properties than
those of (\ref{firga}), (\ref{n31}); this has not been considered in this paper.

\section{Summary\label{sec_sum}}

The new stochastic thermostats presented in this paper are appropriate for
quaternion-based rigid body models. They are written in the form of Langevin
equations and gradient-Langevin system (gradient subsystem for the
translational degrees of freedom and Langevin subsystem for the rotational
degrees of freedom). The obtained stochastic systems preserve the unit
length of the rotational coordinates in the quaternion representation of the
rigid-body dynamics. The thermostats allow to couple both translational and
rotational degrees of freedom to the \textquotedblleft heat
bath\textquotedblright . As it is shown in the numerical tests with the
TIP4P rigid model of water, the Langevin thermostat relaxes to an
equilibrium faster when not only translational degrees of freedom but also
rotational ones are thermostated. It turns out that there is an optimal
range of the strength of coupling to the \textquotedblleft heat
bath\textquotedblright . In contrast, the gradient-Langevin thermostat has a
monotone dependence of relaxation time on the thermostat parameters. In the
case of the Langevin thermostat, two quasi-symplectic second-order (in the
weak sense) integrators are constructed and compared in the numerical tests.
For the gradient-Langevin thermostat, a Runge-Kutta second-order method is
proposed. All the methods preserve the unit length of the rotational
coordinates. The numerical experiments demonstrate the efficiency of the
proposed thermostating technique.

Relaxation times for the gradient-Langevin thermostat are smaller than for
the Langevin thermostat. However, the numerical methods proposed for the
Langevin system have better stability properties than the scheme used for
numerical integration of the gradient-Langevin system. In our experimental
study, the use of the Langevin thermostat together with the quasi-symplectic
integrators was computationally significantly more efficient than
thermostating via the gradient-Langevin system and the numerical scheme for
it.

\begin{acknowledgements}
The work of RLD and RH was supported by the EPSRC research grant
GR/T27105/01 which is gratefully acknowledged. One of the authors
(RLD) did part of the work during his study leave granted by the
University of Leicester.  The computations were performed on the
University of Leicester Mathematical Modelling Centre's cluster,
which was purchased through the EPSRC strategic equipment
initiative.
\end{acknowledgements}




\end{document}